\documentclass{article}
\usepackage[utf8]{inputenc}
\usepackage{geometry}
\pdfoutput=1

\usepackage{auxlib}
\usepackage{authblk}

\title{Logarithmic Approximations for Fair k-Set Selection}
\author[1]{Shi Li}
\author[2]{Chenyang Xu}
\author[3]{Ruilong Zhang}

\affil[1]{\small School of Computer Science, 
Nanjing University, Nanjing, China}
\affil[2]{\small Software Engineering Institute, East China Normal University, Shanghai, China}
\affil[3]{\small Department of Mathematics, Technical University of Munich, Munich, Germany}

\affil[ ]{\texttt{shili@nju.edu.cn, cyxu@sei.ecnu.edu.cn, ruilong.zhang@tum.de}}

\date{}

\newcommand{\opt}{\mathrm{OPT}}
\newcommand{\alg}{\mathrm{ALG}}

\def\FKSS/{\textsc{FKSS}}
\def\WKSS/{\textsc{FKSS}}

\newcommand{\dis}{\mathsf{dis}}

\newcommand{\fX}{\mathfrak{X}}

\newcommand{\dg}{\mathrm{DG}}

\newcommand{\cmu}{\check{\mu}}
\newcommand{\hmu}{\hat{\mu}}

\newcommand{\xiu}{X_u^{(i)}}
\newcommand{\xiiu}{X_u^{(i+1)}}
\newcommand{\xiv}{X_v^{(i)}}
\newcommand{\xiiv}{X_v^{(i+1)}}
\newcommand{\xtu}{X_{u}^{(t)}}
\newcommand{\xip}{X_p^{(i)}}
\newcommand{\xiip}{X_p^{(i+1)}}
\newcommand{\xiq}{X_q^{(i)}}
\newcommand{\xiiq}{X_q^{(i+1)}}

\newcommand{\ph}{\mathsf{p}}
\newcommand{\lf}{\mathsf{L}}
\newcommand{\kid}{\mathsf{kid}}

\newcommand{\lp}{\mathsf{LP}}

\newcommand{\rb}{\texttt{RedBlue}}
\newcommand{\true}{\texttt{true}}
\newcommand{\false}{\texttt{false}}
\newcommand{\flag}{\texttt{flag}}

\usepackage{natbib}

\begin{document}

\maketitle

\begin{abstract}
We study the fair k-set selection problem where we aim to select $k$ sets from a given set system such that the (weighted) occurrence times that each element appears in these $k$ selected sets are balanced, i.e., the maximum (weighted) occurrence times are minimized.
By observing that a set system can be formulated into a bipartite graph $G:=(L\cup R, E)$, our problem is equivalent to selecting $k$ vertices from $R$ such that the maximum total weight of selected neighbors of vertices in $L$ is minimized.
The problem arises in a wide range of applications in various fields, such as machine learning, artificial intelligence, and operations research.

We first prove that the problem is $\NP$-hard even if the maximum degree $\Delta$ of the input bipartite graph is $3$, and the problem is in $\PP$ when $\Delta=2$.
We then show that the problem is also in $\PP$ when the input set system forms a laminar family. 
Based on intuitive linear programming, we show that a dependent rounding algorithm achieves $O(\frac{\log n}{\log \log n})$-approximation on general bipartite graphs, and an independent rounding algorithm achieves $O(\log\Delta)$-approximation on bipartite graphs with a maximum degree $\Delta$.
We demonstrate that our analysis is almost tight by providing a hard instance for this linear programming.
Finally, we extend all our algorithms to the weighted case and prove that all approximations are preserved.

\end{abstract}


\newpage
\section{Introduction}

The problem of fair k-set selection is to select $k$ sets from a given set system such that the (weighted) occurrence times that each element appears in these $k$ selected sets are balanced, i.e., the maximum (weighted) occurrence times among all elements are minimized.
Observe that a set system can be formulated into a bipartite graph $G:=(L\cup R, E)$, i.e., each element and set corresponds to a vertex in $L$ and $R$ respectively; there is an edge between an element vertex and a set vertex if the element is included in the set.
Thus, the fair k-set selection problem is equivalent to finding $k$ vertices from $R$ such that the maximum (total weight of) selected neighbors of vertices in $L$ are minimized.

The above problem aims to balance the frequency of element occurrence within a selected subset of sets. 
Thus, it falls under the umbrella of the subset selection problem, which arises in a wide range of applications in various fields, such as artificial intelligent~\cite{DBLP:conf/nips/DeC22,DBLP:conf/www/MehrotraV23,DBLP:conf/aaai/TschiatschekS017}, machine learning~\cite{DBLP:conf/icml/BoehmerCHMV23,DBLP:conf/nips/LangVS22,DBLP:conf/icml/MirzasoleimanBL20,DBLP:conf/icml/TukanZMRBF23}, data mining~\cite{DBLP:conf/kdd/BaoHZ22,DBLP:conf/kdd/YiWW23}, operations research~\cite{DBLP:journals/ior/HazimehM20,DBLP:journals/ior/MazumderRD23}, to name just a few.
In the following, we present three concrete applications of the problem in the recommended system, feature selection, and facility location.

\begin{itemize}
    \item {\bf Fair Ad Recommendation.} The first application of the above problem is about ad recommendation.
    Fairness plays a vital role in a recommendation systems and an unfair system may harm the benefits of multiple stakeholders~\cite{DBLP:conf/recsys/AbdollahpouriB19,DBLP:conf/kdd/BeutelCDQWWHZHC19,DBLP:journals/tist/LiCXGTLZ23}; see~\cite{DBLP:journals/tois/WangM00M23} for an excellent survey.
    The fair k-set selection problem can model a fair ad recommendation scenario as follows: a recommendation system has $n$ users, and the company needs to select $k$ types of ads from $m$ types to push to these users. 
    A user may dislike some ad types, and this can be learned from her block history. 
    If the user receives too many ads that she does not like to see, this will cause her to resent the recommendation system. 
    We use {\em disagreement} to denote the case where a user sees an ad that she dislikes.
    The goal is to fairly pick $k$ ads from $m$ ads, i.e., minimize the maximum disagreement among all users.
    \item {\bf Fair Feature Selection.} The above problem also finds applications in feature selection.
    Selecting a set of suitable features that ensures these features equally represent all instances in the data set is a crucial problem in machine learning, and this could be particularly useful in situations where the data is unbalanced~\cite{DBLP:conf/sigmod/GalhotraSSV22,DBLP:conf/aistats/QuinzanKHCW0M23}.
    Consider the following scenario where we are working on a dataset for credit card fraud detection, where fraudulent transactions (minority) are significantly fewer than legitimate ones (majority). 
    Each transaction can be considered as an element, and each set is regarded as a set of common features that all transactions in this set share.
    We aim to detect credit card fraud, so some features of a transaction are useless, e.g., the name and age of the trader who makes the transaction. 
    The goal is to select some features such that each transaction, whether fraudulent or legitimate, is equally represented.
    This could result in a model that is more effective at detecting fraudulent transactions despite the imbalance in the data.
    \item {\bf Fair Facility Location.} The last impetus for studying the above problem is an application in facility location. 
    Fairness in facility location has been extensively studied in the field of both game theory and computational social choices~\cite{DBLP:conf/nips/0001LSW22,DBLP:conf/aaai/0001WLC21,DBLP:conf/ijcai/ZhouLC22}.
    Consider the following scenario where the government aims to select $k$ positions from $m$ positions to build facilities (e.g., garbage recycle stations).
    There are $n$ agents, and all agents do not want many garbage recycling stations built near them. 
    Each agent has multiple adjacent positions, which can be viewed as a subset of the given $m$ positions.
    We also use the disagreement to denote that case where some recycling station is built near an agent. 
    The goal is to find a fair way to build $k$ facilities, i.e., minimize the maximum disagreement.
\end{itemize}

Our problem is related to the anonymous refugee housing problem, which was recently proposed by~\cite{DBLP:conf/atal/KnopS23,DBLP:journals/corr/abs-2308-09501}.
In this problem, the input is an undirected graph, and each vertex corresponds to a house that is either an empty house or occupied by an inhabitant.  
There are $k$ anonymous refugees, each of which is required to be assigned an empty house.
Each inhabitant has an upper bound, and an assignment is called inhabitant respected if the number of refugees in the neighborhood of every inhabitant is at most its upper bound.
The goal is to determine whether an inhabitant-respected assignment exists.
By considering (\rom{1}) each inhabitant as an element; (\rom{2}) each empty house as a set of inhabitants that are adjacent to this empty house, the minimax fairness version of the anonymous refugee housing problem is equivalent to our problem.
So, the fair anonymous refugee housing problem can be viewed as another application in Fair Facility Location.
The previous works only focus on the computation complexity while we aim to design approximate algorithms, and this leads to a completely different technique.

For convenience of description, we shall use the bipartite-graph-based formulation of our problem.
By following the convention of the community, we consider each vertex in $L$ as an agent and the integer $k$ as the {\em demand}, where $k$ is the number of vertices that we need to select.
For each vertex in $L$, we use {\em disagreement} to denote the number of selected neighbors of this vertex.
The formal definition of the problem is shown in \cref{sec:model}.

\subsection{Our Contributions}

We consider the Fair k-Set Selection problem (\FKSS/) in which we mainly use the bipartite graph $G:=(L\cup R, E)$ to describe the input.
We distinguish several cases according to the type of the input bipartite graph.
For each case, we give either an exact or approximation algorithm running in polynomial time.
In the following, we summarize the main results of this work.

\paragraph{Main Result 1 (\cref{thm:degree=3:hardness}, \ref{thm:weight:degree=2}, \ref{thm:laminar}).}
The problem \FKSS/ is $\NP$-hard even on bipartite graphs with $\Delta=3$.
For bipartite graphs with $\Delta=2$, the problem \FKSS/ is in $\PP$.
The problem is also in $\PP$ if the input set system forms a laminar family.
Moreover, the maximin criteria do not admit any polynomial time $\alpha$-approximate algorithm unless $\PP=\NP$, where $\alpha$ is an arbitrary function of the input.

Our first result focuses on the computation complexity of the problem (\cref{sec:degree=2}).
We first show that the problem is $\NP$-hard even in the case $\Delta=3$.
The hardness result is built on the maximum independent set on planar graphs, which is shown to be $\NP$-hard in~\cite{DBLP:journals/siamam/GareyJ77}.
We then show that the complement case ($\Delta=2$) is polynomially solvable by giving a simple and efficient combinatorial algorithm.
For the laminar set family, we give a dynamic programming algorithm that computes the optimal solution in polynomial time.
One may expect that maximin criteria are also candidates to investigate the fairness of the k-set selection problem.
Namely, find $k$ sets such that the minimum number of selected neighbors of vertices in $L$ is maximized, where we call these neighbors {\em agreement}.
This is not the case because it is $\NP$-hard to determine whether the optimal solution has a zero agreement; this case is equal to determining whether there exist $k$ sets that cover all elements.

\paragraph{Main Result 2 (\cref{thm:general}, \ref{thm:gap}).}
Given any instance of \FKSS/ on general bipartite graphs, there is a randomized algorithm that achieves $O(\frac{\log n}{\log \log n})$-approximation with high probability running in $\poly(n)$ times, where $n$ is the number of vertices in the graph.
Moreover, our analysis is optimal up to a constant factor.

Our second result focuses on the general bipartite graph (\cref{sec:general}, \ref{sec:weight}).
To have a better understanding of our algorithmic ideas, we focus on the unweighted case in \cref{sec:general} and then we extend our algorithms to the weighted case in \cref{sec:weight}.
We give two algorithms and both algorithms are LP-based rounding algorithms.
The first algorithm is a simple independent rounding algorithm that is $O(\frac{\log n}{\log \log n})$-approximate and it only can satisfy the demand requirement with high probability.
The second algorithm is a dependent rounding algorithm which can surely satisfy the demand requirement while achieving the same ratio.
The random variables used in the second algorithm are negatively correlated, which enables us to use the strong concentration bounds (e.g., Chernoff bound). 
The algorithm is a special case of the classical pipage rounding technique~\cite{DBLP:conf/focs/ChekuriVZ10}, which is used to handle the more general matroid constraints.
We also demonstrate that our analysis is tight up to some constant factor by showing that the LP has a $\Omega(\frac{\log n}{\log \log n})$ integrality gap.

\paragraph{Main Result 3 (\cref{thm:degree=d}).}
Given any instance of \FKSS/ on bipartite graphs with a maximum degree $\Delta$, there is a randomized algorithm that achieves $O(\log \Delta)$-approximation with running time $\poly(n)$ in expectation.
Moreover, our analysis is nearly tight.

By observing that the maximum degree of many practiced graphs in real-life scenarios is often small compared to the number of vertices in the whole graph, our third result focuses on graphs with a maximum degree $\Delta$ (\cref{sec:degree=d}, \ref{sec:weight}).
To have a better understanding of our algorithm, we first focus on the unweighted case in \cref{sec:degree=d} and then extend our algorithms to the weighted case in \cref{sec:weight}.
We prove that there is a randomized algorithm achieving $O(\log\Delta)$-approximation ratio, which significantly improves the $\Delta$ approximation achieved by a trivial algorithm.
Our algorithm is based on the same linear programming formulation as the general graph.
To get rid of the dependency on the number of vertices, we employ the powerful {\em Lovász Local Lemma}.
Our analysis is also almost tight since the same integrality gap instance also implies a $\Omega(\frac{\log\Delta}{\log \log \Delta})$ gap of the natural linear programming.

\subsection{Other Related Works}


Subset selection has numerous variants in literature. 
To the best of our knowledge, \FKSS/ is a novel problem and has never been addressed before.
One of the most representative subset selection problems is to select a subset of elements such that a monotone submodular function is optimized.
If the selected subset is required to satisfy a cardinality constraint, the submodular maximization admits a $(1-\frac{1}{e})$-approximate algorithm~\cite{DBLP:journals/mp/NemhauserWF78}.
This result can be further extended to a more general matroid constraint, and the approximation ratio remains the same~\cite{DBLP:journals/siamcomp/CalinescuCPV11,DBLP:journals/siamcomp/FilmusW14}.
In contrast, the submodular minimization subject to some constraint has a strong lower bound~\cite{DBLP:journals/siamcomp/SvitkinaF11}.
When the selected subset is required to satisfy a knapsack constraint, the submodular maximization also admits a $(1-\frac{1}{e})$-approximate algorithm~\cite{DBLP:journals/orl/Sviridenko04}.


\subsection{Roadmap}

In \cref{sec:model}, we give a formal definition of our problem. 
We show in \cref{sec:degree=2} that our problem is $\NP$-hard even if $\Delta=3$, and we give a simple polynomial-time algorithm that solves the $\Delta=2$ case.
We also show that the problem is in $\PP$ when the input set system forms a laminar family.
In \cref{sec:general}, we consider the problem on general bipartite graphs and give a $O(\frac{\log n}{\log \log n})$-approximate algorithm.
In \cref{sec:degree=d}, we show that the problem admits a $O(\log \Delta)$-approximate algorithm on graphs with a maximum degree $\Delta$.
In \cref{sec:weight}, we extend our algorithms in \cref{sec:general} and \cref{sec:degree=d} to the weighted case.
For a better understanding of our algorithm, we will focus on the cardinality version of our problem when presenting our logarithmic approximation algorithms.

\section{Preliminaries}
\label{sec:model}

We consider the Fair k-Set Selection problem (\FKSS/). 
An instance of \FKSS/ consists of a set system $(U,\cC)$ and a positive integer $k$ with $k\leq \abs{\cC}$, where $k$ is called the {\em demand}.
Let $U:=\set{1,\ldots,n}$ be a set of ground elements and $\cC:=\set{C_1,\ldots,C_m}$ be a collection of subsets defined over the ground element $U$, i.e., $C_j\subseteq U$ for all $j\in[m]$.
Each $C_j\in \cC$ has a non-negative weight $w_j\geq 0$.
A feasible solution $\cS$ is defined as a subcollection of $\cC$ such that $\cS$ contains at least $k$ sets from $\cC$, i.e., $\cS \subseteq \cC$ with $\abs{\cS} \geq k$.
Given a feasible solution $\cS$, the {\em disagreement} of an element $e\in U$, denoted by $\dis_e(\cS)$, is the total weight of sets in $\cS$ that contains $e$, i.e., $\dis_e(\cS):=\sum_{C_j\in \cS:C_j\ni e}w_j$.
The goal is to find a feasible solution $\cS$ such that the maximum disagreement over all elements is minimized, i.e., find a subcollection $\cS\subseteq\cC$ with $\abs{\cS}\geq k$ such that $\max_{e\in[n]}\dis_e(\cS)$ is minimized.

Observe that a set system $(U,\cC)$ can be formulated as a bipartite graph $G:=(L\cup R, E)$.
Formally, we can construct a bipartite graph by the given set system and vice versa: (\rom{1}) for each element $i$ in $U$, we have one vertex $i\in L$; (\rom{2}) for each subset $C_j$ in $\cC$, we have one vertex $j\in R$; (\rom{3}) there is an edge $(i,j)$ in $E$ if and only if the element $i$ is included in the set $C_j$.
Thus, our problem is equivalent to the following subgraph selection problem: 
given a bipartite graph $G:=(L\cup R, E)$ with $\abs{L}=n$ and $\abs{R}=m$, a feasible solution $S\subseteq R$ is a subset of vertices in $R$ with $\abs{S}\geq k$.
Each vertex $v$ in $R$ has a non-negative weight $w_v\geq 0$.
For each vertex $u$ in $G$, let $N_G(u)$ be the set of neighbors of $u$ in $G$.
The disagreement of a vertex $u\in L$ is the total weight of its neighbors that are also in $S$, i.e., $\dis_u(S):=\sum_{v\in N_G(u)\cap S}w_v$.
The goal is to select a subset of vertices $S\subseteq R$ with $\abs{S}\geq k$ such that $\max_{i\in L}\dis_u(S)$ is minimized.

In the remainder of this paper, we will use $\Delta$ to denote the maximum degree of the input bipartite graph $G=(L\cup R,E)$, i.e., $\Delta:=\max_{i\in L\cup R}\abs{N_{G}(i)}$.
In the corresponding set system, $\Delta$ means that each set contains at most $\Delta$ elements, and each element appears in at most $\Delta$ sets.
Without loss of generality, we also assume that the degree of all vertices in $R$ is at least $1$; if not, we just remove such a vertex and decrease the demand by $1$.
This produces an equivalent instance.
Observe that such an assumption implies that any solution has a disagreement at least $1$.

\section{Computation Complexity}
\label{sec:degree=2}

In this section, we mainly show that the problem is $\NP$-hard even on a very restricted graph and all vertices in $R$ have the same weight, i.e., the maximum degree of the input bipartite graph is at least $3$ (\cref{thm:degree=3:hardness}).
This is built on the hardness result of the maximum independent set on planar graphs with a maximum degree of $3$~\cite{DBLP:journals/siamam/GareyJ77}.
For the complement case where the maximum degree is at most $2$, we show that the problem is in $\PP$ by giving a simple algorithm (\cref{thm:weight:degree=2}).
In addition, we prove that the problem is also in $\PP$ when the input set system forms a laminar family (\cref{thm:laminar}).

\subsection{Hardness of Maximum Degree 3}

We shall build the reduction from the problem of the maximum independent set on planar graphs with a maximum degree of $3$.
In~\cite{DBLP:journals/siamam/GareyJ77}, they show that the following deterministic problem is $\NP$-complete (\cref{fact:independent-set}).

\begin{fact}[\cite{DBLP:journals/siamam/GareyJ77}]
Given any planar graph with a maximum degree of $3$, it is $\NP$-complete to determine whether the graph has an independent set of size $p$.
\label{fact:independent-set}
\end{fact}

\begin{theorem}
The problem of \FKSS/ is $\NP$-hard even on the bipartite graph $G:=(L\cup R, E)$ such that (\rom{1}) $\abs{N_G(u)} = 2$ for all $u\in L$; (\rom{2}) $\abs{N_{G}(v)}\leq 3$ for all $v\in R$.   
\label{thm:degree=3:hardness}
\end{theorem} 

To show \cref{thm:degree=3:hardness}, we prove that given any \FKSS/ instance on the desired bipartite graph $G:=(L\cup R,E)$, it is $\NP$-complete to determine whether there exists a solution such that the maximum disagreement of all vertices in $L$ is $1$.

\begin{proof}[Proof of \cref{thm:degree=3:hardness}]
Given any planar graph $G:=(V,E)$ with the maximum degree $3$, we construct a bipartite $B:=(L\cup R, Q)$ as follows: 
(\rom{1}) for each edge $e\in E$, we have one vertex $l$ in $L$; 
(\rom{2}) for each vertex $v\in V$, we have one vertex $r$ in $R$;
(\rom{3}) there is an edge $(l,r)$ between $l\in R$ and $r\in R$ if and only if $r$'s corresponding vertex is one of the endpoint of $l$'s corresponding edge.
Since each vertex in $L$ corresponds to an edge that has exactly two endpoints, we have $\abs{N_{B}(l)}=2$ for all $l\in L$.
Since each vertex in $R$ corresponds to a vertex and the maximum degree of $G$ is $3$, the constructed bipartite graph satisfies $\abs{N_{B}(r)}\leq 3$ for all $r\in R$.
Now, we set the demand $k$ as $p$; recall that, in the independent set problem, we are concerned whether an independent set of size $k$ exists. 

To finish the proof, we only need to prove that if there exists an independent set of size $p$ on $G$, then there exists a solution to \FKSS/ on the constructed bipartite graph such that the maximum disagreement among all vertices in $L$ is $1$, and vice versa.
It is not hard to verify the above claim since a solution to \FKSS/ of maximum disagreement $1$ means that there exist $k$ vertices from $R$ such that they share no common vertex from $L$.
Since each vertex in $L$ corresponds to an edge in the independent set instance, this shall imply that $k$ vertices exist from $V$ such that they share no common edge from $E$.
Hence, an independent set of size $k$ exists and vice versa. 
\end{proof}

\subsection{Algorithms for Maximum Degree 2}

In the section, we show that \WKSS/ is in $\PP$ when the maximum degree of the input bipartite graph is $2$.
Namely, we show the following theorem.
We remark that if all vertices in $R$ have the same weight, there exists a simple linear algorithm for the problem with $\Delta=2$; see \cref{app:sec:delta=2} for details.

\begin{theorem}
The problem \WKSS/ is in $\PP$ on bipartite graphs $G$ with $\Delta\leq 2$.    
\label{thm:weight:degree=2}
\end{theorem}

Note that in the current case, the bipartite graph consists of a set of connected components, each of which is either a cycle or a path.
Consider any feasible solution $\cS$ and any vertex $u\in L$, let $p$ and $q$ be the neighbors of $u$.
Observe that the disagreement $\dis_u(\cS)$ only has $4$ possible values, i.e., $\dis_u(\cS)\in\set{0,w_p,w_q,w_p+w_q}$.
For each $u\in L$, we define $B_u:=\set{0,w_p,w_q,w_p+w_q}$, where $p$ and $q$ are neighbors of $u$.
Let $B:=\bigcup_{u\in L}B_u$ and note that $\abs{B}\leq 4n$ where $n$ is the size of $L$. 
Observe that the optimal solution must be equal to one of the values in $B$.

Before introducing the algorithm, we first present a related problem studied in~\cite{DBLP:journals/corr/abs-2308-09501}.
The problem is stated in \cref{def:weight:degree=2:red-blue}. 

\begin{definition}[Red-Blue]
Given an undirected graph $F:=(R\cup W, D)$ with the maximum degree $2$, where $R\cup W$ is the vertex set and $D$ is the edge set.
Each vertex in $C$ is either red or white, and $R, W$ are the red and white vertex sets, respectively.  
We aim to color $k$ white vertices with blue.
Each red vertex $r$ has an upper bound $\ell_r$ that stands for the maximum number of blue vertices around the red vertex $r$.
The goal is to determine whether there exists a coloring such that (\rom{1}) $k$ white vertices are colored blue; (\rom{2}) for each red vertex $r$, the number of blue vertices is at most $\ell_r$.    
\label{def:weight:degree=2:red-blue}
\end{definition}

In~\cite[Theorem 3]{DBLP:journals/corr/abs-2308-09501}, they show that the Red-Blue problem can solved efficiently; see \cref{lem:weight:degree=2:red-blue}.

\begin{lemma}[\cite{DBLP:journals/corr/abs-2308-09501}]
The Red-Blue problem stated in \cref{def:weight:degree=2:red-blue} can be determined in $O(n)$ times, where $n$ is the number of vertices in the input graph.
\label{lem:weight:degree=2:red-blue}
\end{lemma}

Given any undirected graph $F:=(R\cup W, D)$, let $\rb(\cdot,\cdot,\cdot,\cdot)$ be the algorithm that is able to determine whether the instance admits a desired coloring.
The algorithm $\rb(R,W,k,\bu)$ takes four parameters as the input: $R$ is a set of red vertices, $W$ is a set of white vertices, $k$ is the number of white vertices that are required be colored with blue, and $\bu:=(\ell_u)_{u\in R}$ is the upper bound vector.
Taking these four parameters as input, the algorithm returns either $\true$ or $\false$, which indicates whether the instance admits a feasible coloring, i.e., $\rb(R,W,k,\bu)\in\set{\true,\false}$.

Our algorithm is built based on $\rb(\cdot,\cdot,\cdot,\cdot)$ by observing that the red vertex corresponds to a vertex in $L$ and the white vertex corresponds to a vertex in $R$ in our problem.
Note that the optimal solution only has $4n$ possible values.
To this end, given any value $b$ from $B$, we only need to determine the upper bound vector so that $\rb(\cdot,\cdot,\cdot,\cdot)$ can be used to determine whether there exists a feasible solution under the value of $b$. 
Since each vertex in $L$ only has two neighbors, there are only three cases, and thus, the upper bound is easy to compute; see lines~\ref{line:weight:degree=2:start}-\ref{line:weight:degree=2:end} of \cref{alg:weight:degree=2} for details.

\begin{algorithm}[htb] 
\caption{Algorithm for \WKSS/ with $\Delta\leq 2$}
\label{alg:weight:degree=2}
\begin{algorithmic}[1]
\Require The bipartite graph $G:=(L\cup R, E)$; The demand $k$; The algorithm $\rb(\cdot,\cdot,\cdot,\cdot)$; The value set $B$.
\Ensure A vertex set $S$ with $\abs{S}=k$.
\State Sort all values in $B$ in non-decreasing order.
\Comment{$B:=\set{b_1,\ldots,b_{\kappa}}$}
\State $\flag\leftarrow\false$; $i\leftarrow 0$.
\While{$\flag=\false$}
\State $i\leftarrow i+1$.
\Comment{Try next value}
\For{each vertex $u\in L$}
\Comment{$B_u:=\set{0,w_p,w_q,w_p+w_q}$}
\IIf{$b_i\geq w_p+w_q$} $\ell_u \leftarrow 2$; \Break, \EndIIf
\label{line:weight:degree=2:start}
\IIf{$b_i\geq w_p$ or $b_i\geq w_q$} $\ell_u \leftarrow 1$; \Break, \EndIIf
\IIf{$b_i< w_p$ and $b_i<w_q$} $\ell_u\leftarrow 0$; \Break, \EndIIf
\label{line:weight:degree=2:end}
\EndFor
\State $\flag \leftarrow \rb(L,R,k,(\ell_u)_{u\in L})$.
\EndWhile
\State Let $\mathfrak{C}$ be the coloring scheme of $\rb(\cdot,\cdot,\cdot,\cdot)$ by using the last round of $(\ell_u)_{u\in L}$.
\State \Return $S\leftarrow\set{v\in R\mid v \text{ is colored with blue in }\mathfrak{C}}$.
\end{algorithmic}
\end{algorithm} 

\begin{proof}[Proof of \cref{thm:weight:degree=2}]
Note that the optimal solution must be one of the values in $B$, and the size of $B$ is $\poly(n)$.
In \cref{alg:weight:degree=2}, we try every value in $B$ from small to large.
For each value $b$, we use $\rb(\cdot,\cdot,\cdot,\cdot)$ to determine whether the instance admits a feasible solution.
Since $\rb(\cdot,\cdot,\cdot,\cdot)$ runs in polynomial time, we find the optimal solution in $\poly(n)$ times.
\end{proof}

\subsection{Algorithms for Laminar Instance}
\label{subsec:laminar}

In this section, we consider a special case of \FKSS/ where the set system is a {\em laminar family}.
The laminar family captures the hierarchical structure that widely appears in real-world applications.
The laminar family defines a collection of uncrossed sets.
Formally, a set system $(U,\cC)$ is called a laminar family if, for every two distinct sets $W,Q\in \cC$, only one of the following statement holds: (\rom{1}) $W\subsetneq Q$, or; (\rom{2}) $Q\subsetneq W$, or; (\rom{3}) $W\cap Q=\emptyset$.
An example can be found in \cref{fig:gap_instance}.
We aim to show the following theorem (\cref{thm:laminar}) in this section.

\begin{theorem}
The problem of \FKSS/ can be solved in polynomial time if the input set system is a laminar family.
\label{thm:laminar}
\end{theorem}

\begin{figure}[htb]
    \centering
    \includegraphics[width=1\linewidth]{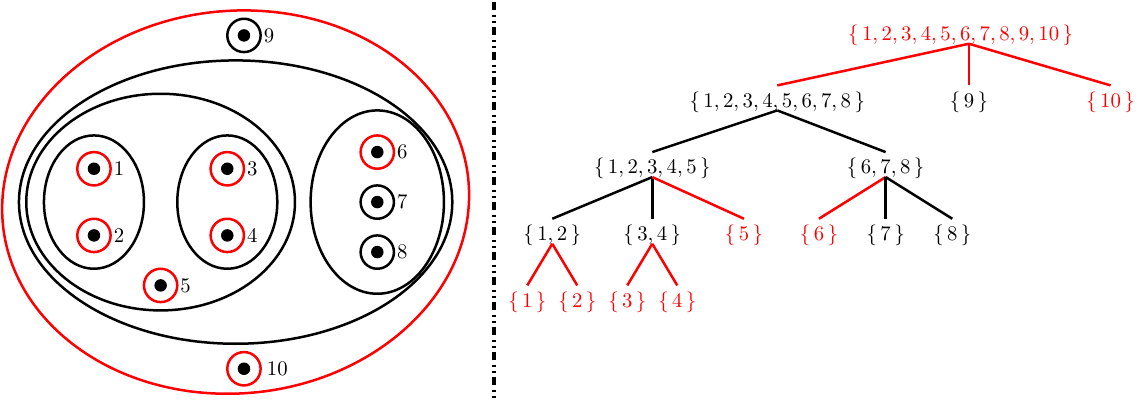}
    \caption{An example of a laminar family and its corresponding tree. The original laminar family is shown on the left and marked as black lines. We add some dummy sets (whole ground element sets and all singleton sets), and they are marked as red lines. The corresponding tree is shown on the right, where black lines and sets are from the original instance, and red lines and sets are constructed dummy sets.}
    \label{fig:laminar}
\end{figure}

Given a laminar family $(U,\cL)$, the collection of subsets $\cL$ can be naturally represented as a forest structure.
Formally, for every subset $C\in\cL$, we create a node $\bv$.
A set $W\in\cL$ is a parent set of a set $Q\in\cL$ if and only if (\rom{1}) $Q\subseteq W$; (\rom{2}) there is no set $P$ such that $Q\subseteq P \subseteq W$.  
By adding some dummy sets, the forest structure can be modified as a tree structure, i.e., (\rom{1}) add the whole ground set $U$ to $\cL$ if $U\notin \cL$; (2) add all singleton sets $\set{e}$ to $\cL$ if $\set{e}\notin \cL$.
An example can be found in \cref{fig:laminar}.

Given any laminar family $(U,\cL)$, it is not hard to see that such a tree structure $\bT:=(\bV,\bE)$ can be constructed in linear time.
Thus, \FKSS/ on the laminar family can be reduced to the following node-selection problem on the tree.

\paragraph{Node-Selection-on-Tree.}
Given a tree $T:=(V,E)$ with root node $r$, every node $v\in V$ has a non-negative weight $w_v\geq 0$.
We use $\lf\subseteq V$ to denote the set of leaf nodes in $T$.
Given any node set $V'\subseteq V$, we use $w(V')$ to denote the total weight of nodes in $V'$.
For any node $v\in V$, let $\ph(v)\subseteq V$ be a set of nodes that are included in the path from the root $r$ to $v$.
Note that $\ph(v)$ is unique for any $v\in V$.
Given any node set $V'$, the disagreement of a leaf node $u$ is defined as the total weight of nodes in both $V'$ and $\ph(u)$, i.e., $\dis(u,V'):=\sum_{v\in V'\cap\ph(u)}w_v$.
The goal is to select a node set $V'$ with $\abs{V'}=k$ such that the maximum disagreement in $\lf$ is minimized, i.e., $\max_{v\in\lf} \dis(v,V')$ is minimized.

In the proof of \cref{lem:laminar}, we show that the node selection problem can be solved by a dynamic programming.
The dynamic programming works mainly due to a simple but crucial property on the tree: when some node $u$ is added to the solution, the disagreement of all leaf nodes that are in the subtree rooted at $u$ will increase by $u$'s weight.
Thus, this property implies that the problem has some optimal structure which enables a dynamic programming method.
\begin{lemma}
The problem of node selection on a tree can be solved in polynomial time.
\label{lem:laminar}
\end{lemma}

\begin{proof}
We create a DP table as follows: for each vertex $u\in V$, let $D(u,x)$ store the optimal solution to the subinstance $(T',x)$, i.e., $D(u,x)$ is the minimum value of max-disagreement among all leaf nodes in the subtree rooted at $u$ (denoted by $T'$), where the algorithm is required to select $x$ nodes from $T'$.
The range of $x$ is $\set{0,1,\ldots,k}$.
Let $\kid(u)$ be a set of child nodes of $u$.
Then, $D(u,x)$ can be computed by $D(v,x),v\in\kid(u)$.
There are two cases: (\rom{1}) $u$ is selected; (\rom{2}) $u$ is not selected.
In the first case, we have $D(u,x)=w_u+\min_{v\in\kid(u)}D(v,x-1)$.
In the second case, we have $D(u,x)=\min_{v\in\kid(u)}D(v,x)$.
In summary, the formula of $D(u,x)$ is:
$$
D(u,x)=
\min\left\{ w_u+\min_{v\in\kid(u)}D(v,x-1), \min_{v\in\kid(u)}D(v,x) \right\}.
$$
The total number of sub-problems that we have is at most $n\cdot (k+1)$, where $n$ is the number of nodes in the tree and $k\leq n$ is the demand.
For each subproblem, computing the optimal solution requires $O(n)$ times.
Thus, the running time of the DP above is $O(n^3)$.
\end{proof}

\paragraph{Wrapping Up.} 
Now, we are ready to prove \cref{thm:laminar}.

\begin{proof}[Proof of \cref{thm:laminar}]
Given any laminar family $(U,\cL)$, we add two types of dummy sets to $\cL$, i.e., the whole ground element sets and all singleton sets.
This leads to a new laminar family $(U,\cL')$.
Note that in $\cL'$, all elements have a singleton set, which may be either an original set or a dummy set.
We construct a tree structure $\bT:=(\bV,\bE)$ for $(U,\cL')$ and this can be done in linear time.
Note that $\bT$ consists of two types nodes: (\rom{1}) nodes corresponds a set in $\cL$; (\rom{2}) nodes corresponds a dummy set in $\cL'$.  
For each set $C$ in $\cL$, we have one node $\bv_{C}$ with weight $w_C$.
For each dummy set $F$, we have one node $\bv_{F}$ with weight $w_F=+\infty$.

In the constructed tree $\bT$, each leaf node corresponds to a singleton set, and all singleton sets in $\cL'$ corresponds to a leaf node.
Now, we run the DP algorithm stated in the proof of \cref{lem:laminar} to find the optimal solution (denoted by $\bV'$) to the node selection problem on $\bT$.
Note that $\abs{\bV'}=k$, and we choose the corresponding set, which forms a feasible solution $\cS$ to \FKSS/.
Note that $\cS\subseteq \cL$ must hold since the dummy set has a very large weight.
Since each leaf node is a singleton set, the disagreement of each leaf node in $\bT$ is exactly the same as the disagreement of each element in $(U,\cL')$.
Thus, $\cS$ is an optimal solution to \FKSS/ due to the optimality of $\bV'$.
\end{proof}

\section{General Graphs}
\label{sec:general}

In this section, we give two logarithmic approximate algorithms.
To better understand our algorithmic idea, we focus on the case where all vertices in $R$ have the same weight.
We will extend our algorithms to the general \FKSS/ instance in \cref{sec:weight}.
Both algorithms are LP-based randomized rounding algorithms.
The first algorithm is a simple independent rounding algorithm that (\rom{1}) achieves the $O(\frac{\log n}{\log \log n})$-approximation with high probability; (\rom{2}) satisfies the demand requirement with high probability; see \cref{subsec:independent} for details.
The second algorithm is a dependent rounding algorithm which strictly improves the first algorithm, i.e., it (\rom{1}) achieves $O(\frac{\log n}{\log \log n})$-approximation with high probability; (\rom{2}) satisfies the demand requirement with probability $1$; see \cref{subsec:pipage}.
The dependent rounding algorithm is a special pipage rounding algorithm proposed by~\cite{DBLP:conf/focs/ChekuriVZ10}, which preserves the negative correlation property.
The main advantage of the independent rounding algorithm is the simple description and analysis of the algorithm.
The analysis of special pipage rounding requires the proof of the negative correlation property proved by~\cite{DBLP:conf/focs/ChekuriVZ10}.

Subsequently, we show that our analysis is optimal up to the constant factor by giving a hard instance (\cref{thm:gap}), i.e., the linear programming formulation that we used has an integrality gap of $\Omega(\frac{\log n}{\log \log n})$.

\subsection{LP Formulation}

As is typical, we shall start from the following intuitive linear programming formulation.
For each vertex $v$ in $R$, we have a variable $x_v$, where $x_v=1$ means that we choose vertex $v$; $x_v=0$ otherwise.
The first constraint ensures that the disagreement of all vertices in $L$ is at most $T$.
The second constraint ensures that the number of selected vertices satisfies the required demand.

\begin{align*}
    \text{min} && T & \tag{\text{Intu-LP}} \label{Intu-LP}\\
    \text{s.t.} \nonumber \\
    &&\sum_{v\in N_G(u)}x_v &\leq T, &\forall u\in L \\
    &&\sum_{v\in R}x_v &\geq k, & \\
    && 0\leq x_v &\leq 1, &\forall v\in R 
\end{align*}

Not surprisingly, \eqref{Intu-LP} has a bad integrality gap, which can be as large as $\Omega(n)$, where $n$ is the number of vertices in the input bipartite graph.
The gap instance is simple, where the input bipartite graph is just a matching (i.e., the bipartite graph consists of $n$ disjoint edges), and the demand $k$ is $1$.
In this instance, any optimal integral solution must be $1$.
But, the optimal fractional solution shall equally distribute this unit demand over all vertices in $R$ (e.g., $x_v=\frac{1}{n}$ for all $v\in R$); thus, the optimal fractional solution has a maximum disagreement of $\frac{1}{n}$.
This leads to a $\Omega(n)$ gap.

To eliminate the above bad instance, one common technique is to guess the value of the optimal solution by removing the objective of \eqref{Intu-LP}.
This produces the following feasibility-checking linear programming \eqref{Feas-LP}.

\begin{align*}
     &&  & \tag{\text{Feas-LP}} \label{Feas-LP}\\
    &&\sum_{v\in N_G(u)}x_v &\leq T, &\forall u\in L \\
    &&\sum_{v\in R}x_v &\geq k, & \\
    && 0\leq x_v &\leq 1, &\forall v\in R 
\end{align*}

We shall guess the value of the optimal solution whose range is from $1$ to $k$. 
Hence, \eqref{Feas-LP} eliminates the bad instance of \eqref{Intu-LP} because the guessed value $T\geq 1$.
Let $T^*$ be the minimum integer in $[1,k]$ to make \eqref{Feas-LP} admit a feasible solution; $T^*$ can be obtained by solving $O(\log k)$ times \eqref{Feas-LP}.
Observe that $T^*$ is a lower bound of the optimal solution, i.e., any optimal integral solution has a maximum disagreement of at least $T^*$.
Let $\bx^*:=(x_v^*)_{v\in R}$ be the solution to \eqref{Feas-LP} that achieves $T^*$.

\subsection{Independent Rounding}
\label{subsec:independent}

In this section, we show that there is a simple independent rounding algorithm that (\rom{1}) achieves $O(\log n)$-approximation with high probability; (\rom{2}) satisfies the demand requirement with high probability.
Formally, we aim to prove the following theorem (\cref{thm:independent}).

\begin{theorem}
There is a randomized rounding algorithm that (\rom{1}) achieves $O(\log n)$-approximation with probability at least $1-\frac{1}{n}$; (\rom{2}) satisfies the demand requirement with probability at least $1-O(1/\frac{\ln n}{\ln \ln n})$, where $n$ is the number of vertices in the input bipartite graph.
\label{thm:independent}
\end{theorem}

\paragraph{Algorithmic Intuition.}
The algorithmic idea of \cref{alg:independent} is simple.
We first solve \eqref{Feas-LP} to obtain a fractional solution $(x_v^*)_{v\in R}$, and then, we interpret $x_v^*$ as the probability that we choose $v$.
To ensure that we can sample a sufficient number of vertices from $R$, we raise the probability that we choose $v$ to $O(\frac{\ln n}{\ln \ln n}) \cdot x^*_v$.
Then, by using the Chebyshev bound, we can show that the sampled solution satisfies the demand requirement with high probability.
Meanwhile, the disagreement of each vertex in $L$ will also be raised to $O(\frac{\ln n}{\ln \ln n})\cdot T^*$, where $T^*$ is a lower bound of the optimal solution.
By using the right tail Chernoff bound and union bound, we can show that the disagreement of all vertices in $L$ is at most $O(\frac{\ln n}{\ln \ln n})\cdot\opt$ with high probability.

\begin{algorithm}[htb] 
\caption{Independent Rounding Algorithm.}
\label{alg:independent}
\begin{algorithmic}[1]
\Require The fractional solution $\bx^*$.
\Ensure A set of vertices $S\subseteq R$ with $\abs{S}\geq k$ with high probability.
\State $A\leftarrow \set{v\in R: x_v^*\geq \frac{\ln \ln n}{10 \ln n}}$; $B\leftarrow \set{v\in R: x_v^* < \frac{\ln \ln n}{10 \ln n}}$; $B'\leftarrow \emptyset$.
\If{$\sum_{v\in B} x_v^* \leq 1$} 
\State $B'\leftarrow \set{u}$; $u$ is an arbitrary vertex in $B$.
\EndIf
\If{$\sum_{v\in B} x_v^* > 1$}
\For{each $v\in B$}
\State $p_v\leftarrow \frac{10 \ln n}{\ln \ln n}\cdot x_v^*$.
\Comment{Note that $p_v\leq 1$.}
\State Independently add $v$ to $B'$ with probability $p_v$.
\EndFor
\EndIf
\State \Return $S\leftarrow A\cup B'$.
\end{algorithmic}
\end{algorithm} 

In the following, we show that the returned solution of \cref{alg:independent} satisfies the demand requirement (\cref{lem:independent:size}) and approximation-ratio requirement (\cref{lem:independent:load}) with high probability.
In the following proof, we will use the same notations stated in \cref{alg:independent}.

\begin{lemma}
\cref{alg:independent} returns a vertex set that satisfies the demand requirement with probability at least $1-O(1/\frac{\ln n}{\ln \ln n})$, where $n$ is the number of vertices.
\label{lem:independent:size}
\end{lemma}

\begin{proof}
There are two cases: (\rom{1}) $\sum_{v\in B}x^*_v \leq 1$; (\rom{2}) $\sum_{v\in B}x^*_v >1$.
In the first case, $S=A\cup\set{u}$ for some $u\in B$.
Thus, we have $\abs{S}=1+\abs{B}\geq \sum_{v\in B}x_v^*+\sum_{v\in A}x_v^*=k$; thus, the lemma holds.
In the second case, observe that if we have $\abs{B'}\geq \sum_{v\in B}x^*_v$ with high probability, then $\abs{S}=\abs{A}+\abs{B'}\geq \sum_{v\in A}x^*_v+\sum_{v\in B}x^*_v =k $; thus, the lemma holds with high probability.

In the following, we aim to show $\abs{B'}\geq \sum_{v\in B}x^*_v$ holds with high probability.
For each vertex $v\in B$, define a random variable $X_v$: $X_v=1$ implies $v\in B'$; otherwise, $v\notin B'$.
Thus, $X_v=1$ with probability $p_v=10x_v^*\frac{\ln n}{\ln \ln n}$ and $\abs{B'}=\sum_{v\in B}X_v$.
Note that $p_v\leq 1$ for all $v\in B$ by the definition of the vertex set $B$.
Hence, we aim to show $\sum_{v\in B}X_v \geq \sum_{v\in B}x^*_v$ holds with high probability.
We define $X:=\sum_{v\in B}X_v$ and observe that 
$$
\E[X]=\sum_{v\in B}p_v = \sum_{v\in B}\frac{10 \ln n}{\ln \ln n} \cdot x_v^*=\frac{10\ln n}{\ln \ln n}\cdot\sum_{v\in B}x^*_v \geq \frac{10 \ln n}{\ln \ln n}, 
$$
where the last inequality is due to the condition of Case (\rom{2}): $\sum_{v\in B}x_v^*\geq 1$.
Note that we aim to prove that $X<\sum_{v\in B}x_v^*$ happens with low probability, i.e.,
$$
\Pr[X<\sum_{v\in B}x_v^*]\leq \Pr[\abs{X-\E[X]}>\E[X]-\sum_{v\in B}x_v^*] = \Pr\left[\abs{X-\E[X]}>\frac{\E[X]-\sum_{v\in B}x_v^*}{\sqrt{\E[X]}}\cdot\sqrt{\E[X]}\right].
$$
Note that the variance of a sum of independent random variables is the sum of the variances.
Thus, we have $\V[X]=\sum_{v\in B}p_v\cdot (1-p_v)$; so $\V[x]\leq \E[X]$ since $(1-p_v)\leq 1$ for all $v\in B$.
We will use the Chebyshev bound: 
$$
\Pr[\abs{X-\E[X]}\geq k \cdot \sqrt{\V[X]}] \leq \frac{1}{k^2},\forall k>0.
$$
Thus, we have:
\begin{align*}
\Pr[X<\sum_{v\in B}x_v^*]
&\leq \Pr\left[\abs{X-\E[X]}>\frac{\E[X]-\sum_{v\in B}x_v^*}{\sqrt{\E[X]}}\cdot\sqrt{\E[X]}\right]  \tag*{[Inequality above]} \\
&\leq \Pr\left[\abs{X-\E[X]}>\frac{\E[X]-\sum_{v\in B}x_v^*}{\sqrt{\E[X]}}\cdot\sqrt{\V[X]}\right]  \tag*{[$\V[x]\leq \E[X]$]} \\
&\leq \left( \frac{\sqrt{\E[X]}}{\E[X]-\sum_{v\in B}x_v^*} \right)^2 \tag*{[Chebyshev bound]} \\
&\leq \frac{\ln \ln n}{10\ln n- \ln \ln n} \tag*{[$\E[X]\geq \frac{10\ln n}{\ln \ln n}$]}
\end{align*}
Thus, $\abs{B'}\geq \sum_{v\in B}x^*_v$ holds with probability at least $1-O(1/\frac{\ln n}{\ln \ln n})$.
By repeating the experiment, the probability can be further improved.
\end{proof}

\begin{lemma}
$\Pr[\abs{N_G(u)\cap B'} \geq \frac{10\ln n}{\ln \ln n} \cdot T^*] \leq \frac{1}{n^2}$ for all vertices $v\in L$.
\label{lem:independent:load}
\end{lemma}

\begin{proof}
We consider an arbitrary vertex $u\in L$, and for each vertex $v\in N_G(u)\cap B$, define $X_v$ as a random variable, where $X_v=1$ implies $v$ is in $B'$; otherwise, $v$ is not in $B'$.
If $\sum_{v\in B}x^*_v \leq 1$, then $\abs{B'}=1$; thus, $\abs{N_G(u)\cap B'}\leq 1 \leq \frac{10\ln n}{\ln \ln n} T^* $ with probability 1.
Hence, we focus on the case where $\sum_{v\in B}x^*_v >1$.
Note that $\Pr[X_v=1]=\frac{10\ln n}{\ln \ln n}\cdot x_v^*$.
Define $X:=\sum_{v\in N_G(u)\cap B}X_v$; thus, $X=\abs{N_G(u)\cap B'}$.
We aim to prove that $\Pr[X\geq \frac{10\ln n}{\ln \ln n}\cdot T^*]\leq \frac{1}{n^2}$.
We have the following upper bound on $\E[X]$: 
$$
\E[X]=\sum_{v\in N_G(u)\cap B}\Pr[X_v=1]=\frac{10\ln n}{\ln \ln n}\sum_{v\in N_G(u)\cap B}x^*_v \leq \frac{10\ln n}{\ln \ln n} \cdot T^*.
$$
Now, we apply the right tail Chernoff bound: for all $\delta\geq 1$ and $\E[X]\leq \hat{\mu}$.
$$
\Pr[X\geq (1+\delta) \cdot \hat{\mu}] \leq \exp{\left( -\hmu\cdot (1+\delta)\cdot \ln(1+\delta)/4\right)}.
$$
Applying the Chernoff bound above and setting $1+\delta=\frac{10\ln n}{\ln \ln n}$, we have
\begin{align*}
&\Pr[X\geq (1+\delta)\cdot T^*] \leq e^{-T^*(1+\delta)\ln(1+\delta)/4}  \tag{(Chernoff bound above)} \\
\implies &\Pr[X\geq \frac{10\ln n}{\ln \ln n}\cdot T^*] \leq e^{-T^*2\ln n} \tag{$\frac{10\ln n}{\ln\ln n}\ln\left( \frac{10 \ln n}{\ln \ln n} \right)\geq 8\ln n$} \\
\implies & \Pr[X\geq \frac{10\ln n}{\ln \ln n}\cdot T^*] \leq e^{-2\ln n} = \frac{1}{n^2} \tag{$T^*\geq 1$}
\end{align*}
\end{proof}

\paragraph{Wrapping Up.} Now, we are ready to prove \cref{thm:independent}.

\begin{proof}[Proof of \cref{thm:independent}]
Note that the returned solution of \cref{alg:independent} is $A\cup B'$.
We consider an arbitrary vertex $u\in L$, and observe if we can prove that $\abs{N_G(u)\cap A} \leq \frac{10\ln n}{\ln \ln n}\cdot T^*$ with probability 1, then by \cref{lem:independent:load} we have 
\begin{align*}
\Pr[\abs{N_G(u)\cap S} \geq \frac{20\ln n}{\ln \ln n}\cdot T^*] 
&= \Pr[\abs{N_G(u)\cap A} + \abs{N_G(u)\cap B'} \geq \frac{10\ln n}{\ln \ln n}\cdot T^*] \\
&= \Pr[\abs{N_G(u)\cap B'} \geq \frac{10\ln n}{\ln \ln n}\cdot T^*] \\
&\leq \frac{1}{n^2}.
\end{align*}
$\abs{N_G(u) \cap A} \leq \frac{10\ln n}{\ln \ln n}\cdot T^*$ holds because of the following inequalities: (\rom{1}) $\sum_{v\in N_G(u)\cap A}x_v^* \leq T^*$; (\rom{2}) $1\geq x^*_v\geq \frac{\ln \ln n}{10\ln n}$ for all $v\in N_G(u)\cap A$.
Thus, by union bound, we have 
$$
\Pr[\abs{N_G(u)\cap S}\leq \frac{20\ln n}{\ln \ln n}\cdot T^*\leq \frac{20\ln n}{\ln \ln n}\cdot \opt] \geq 1-\frac{1}{n}.
$$
Hence, \cref{alg:independent} returns a solution that (\rom{1}) achieves $(\frac{20\ln n}{\ln \ln n})$-approximation with probability at least $1-\frac{1}{n}$; (\rom{2}) satisfies the demand requirement with probability at least $1-O(1/\frac{\ln n}{\ln \ln n})$.
\end{proof}

\subsection{Dependent Rounding}
\label{subsec:pipage}

In this section, we present a special pipage rounding and mainly show the following theorem.

\begin{theorem}
Given any instance of \FKSS/ on general bipartite graphs, there is a randomized algorithm with running time $\poly(n)$ that returns a $O(\frac{\log n}{\log \log n})$-approximate solution with probability at least $1-\frac{1}{n}$, where $n$ is the number of vertices in the input bipartite graph. 
\label{thm:general}
\end{theorem}

As we have seen in \cref{subsec:independent}, independent rounding allows us to use the strong concentration bounds but it may not be able to guarantee that the total number of selected vertices satisfies the demand. 
Fortunately, several dependent rounding techniques can ensure the {\em negative correlation} property with which the strong concentration bounds are still applicable.
One of the most elegant methods is called {\em pipage rounding} proposed by~\cite{DBLP:conf/focs/ChekuriVZ10}.
The initial pipage rounding technique in ~\cite{DBLP:conf/focs/ChekuriVZ10} is used to handle the matroid structure.
However, in our problem, the constraint can be viewed as a simple uniform matroid.
Thus, we only need to use a special case of the classical pipage rounding technique, whose descriptions are considerably simpler than the general pipage rounding technique.
In the following, we present the special pipage rounding algorithm in \cref{alg:pipage}.

\paragraph{Algorithmic Intuition.}
The algorithm runs in rounds and, in each iteration, it starts from a feasible fractional solution $\bx$ and picks two fractional variables of $\bx$, denoted by $x_u$ and $x_v$ ($0<x_u,x_v<1$).
Observe that these two variables must exist since $\bx$ is not integral, and the summation of all variables in $\bx$ is an integer.
The algorithm aims to make $x_u$ or $x_v$ be an integer while maintaining the feasibility of the solution (i.e., ensuring the summation of all variables is still $k$).
Hence, the algorithm has two directions to modify $x_u$ and $x_v$: (D-1) decreases $x_u$ and increases $x_v$; (D-2) increases $x_u$ and decreases $x_v$.
Regardless of the direction, the algorithm modifies $x_u$ and $x_v$ at the same rate such that either $x_u$ or $x_v$ becomes an integer ($0$ or 1).
Assume that $\bx_1$ and $\bx_2$ are the solutions produced by the algorithm from $\bx$ via (D-1) and (D-2), respectively.
Note that both $\bx_1$ and $\bx_2$ have a strictly larger number of integral variables than $\bx$.
Instead of checking which direction is profitable, the algorithm chooses a direction randomly, i.e., with probability $p$, $\bx$ is rounded to $\bx_1$; with probability $1-p$, the $\bx$ is rounded to $\bx_2$, where $p$ is a chosen probability in this iteration.
The choice of $p$ (line~\ref{line:pipage:prob} of \cref{alg:pipage}) is crucial to both the marginal probability preservation and the negative correlation property, which will be clear in the proof of \cref{lem:general:pipage:expectation}.

\begin{algorithm}[htb] 
\caption{Special Pipage Rounding.}
\label{alg:pipage}
\begin{algorithmic}[1]
\Require The fractional solution $\bx^*$.
\Ensure A set of vertices $S\subseteq R$ with $\abs{S}\geq k$ with probability $1$.
\While{there exists a variable $x_v^*$ such that $0<x^*_v<1$}
\State Pick two fractional variables $x_v^*$ and $x_u^*$.
\State $\delta_1\leftarrow \min\{ x^*_u, 1-x^*_v \}$.
\State $\delta_2\leftarrow \min\{ 1-x^*_u , x^*_v \}$.
\State $p\leftarrow \frac{\delta_2}{\delta_1+\delta_2}$.
\label{line:pipage:prob}
\State With probability $p$: $x^*_u\leftarrow x^*_u-\delta_1$; $x^*_v\leftarrow x^*_v+\delta_1$.
\Comment{(D-1)}
\State \hspace{2.13cm} Else: $x^*_u\leftarrow x^*_u+\delta_2$; $x^*_v\leftarrow x^*_v-\delta_2$.
\Comment{(D-2)}
\EndWhile
\State \Return $S\leftarrow\set{v\in R \mid x^*_v=1}$.
\end{algorithmic}
\end{algorithm} 

\begin{observation}
\cref{alg:pipage} returns a feasible integral solution in $O(k)$ times, where $k$ is the demand.
\label{obs:pipage:running-time}
\end{observation}


It is shown in~\cite{DBLP:conf/focs/ChekuriVZ10} that \cref{alg:pipage} can produce a collection of negatively correlated variables, which enables us to use the Chernoff bound.
We start with the definition of the negative correlation and the Chernoff bound for negatively correlated variables.

\begin{definition}[Negative Correlation]
Let $\set{X_1,\ldots,X_n}$ be a collection of random variables. 
This collection of random variables is {\em negatively correlated} if and only if for any $S\subseteq [n]$:
$$
\E\left[\prod_{i\in S} X_i \right] \leq \prod_{i\in S}\E[X_i]
$$
\label{def:negative-correlation}
\end{definition}

The classical Chernoff bound requires all variables to be independent.
However, it is also known that the Chernoff bound also works for variables with negative correlation properties.

\begin{definition}[Chernoff Bound with Negative Correlation]
Let $\set{X_1,\ldots,X_n}$ be a collection of random variables that are negatively correlated, where $X_i\in[0,1]$.
Define $X:=\sum_{i\in[n]}X_i$ and $\mu:=\sum_{i\in [n]}\E[X_i]$.
Let $\hmu\geq \mu$ be an upper bound of the expectation.
For any $\delta\geq 1$, we have
$$
\Pr[X\geq (1+\delta) \cdot \hmu ] \leq e^{-\hmu\cdot (1+\delta)\cdot \ln(1+\delta)/4}.
$$
\label{def:chernoff}
\end{definition}

For each vertex $v\in R$, define a random variable $X_v$: $X_v=1$ implies that $v$ is selected by \cref{alg:pipage}; otherwise $v$ is not selected by the algorithm.
In~\cite{DBLP:conf/focs/ChekuriVZ10}, they have shown that \cref{alg:pipage} has the following properties (\cref{lem:general:pipage:pros}).
We restate the proof of \cref{lem:general:pipage:pros} in \cref{app:lem:pipage} for completeness.

\begin{lemma}[\cite{DBLP:conf/focs/ChekuriVZ10}]
For any vertex $v\in R$, we have $\Pr[v\in S]=x_v^*$. Moreover, the collection $\set{X_v}_{v\in R}$ of random variables is negatively correlated.
\label{lem:general:pipage:pros}
\end{lemma}

\begin{lemma}
For each $u\in L$, $\Pr \left[ \abs{N_{G}(u) \cap S } \geq \frac{10 \ln n}{\ln \ln n} \cdot T^* \right] \leq \frac{1}{n^2}$.   
\label{lem:general:ratio}
\end{lemma}

\begin{proof}
The proof is standard and we state only for completeness.
We consider an arbitrary vertex $u\in L$, and define $X:=\sum_{v\in N_G(u)}X_v$.
Note that $X=\abs{N_G(u)\cap S}$ by the definition of random variables.
By \cref{lem:general:pipage:pros}, we have $\Pr[u\in S]=x^*_u$ for all $u\in R$; this is equivalent to $\Pr[X_u=1]=x^*_u$ for all $u\in R$.
Thus, we have the following upper bound of the expectation of the random variable $X$: $\E[X]=\sum_{v\in N_G(u)}x^*_v \leq T^*$.
By \cref{lem:general:pipage:pros}, the random variable collection $\set{X_v}_{v\in R}$ is negatively correlated; thus, we are able to use the Chernoff bound stated in \cref{def:chernoff}.
By choosing $(1+\delta)=\frac{10\ln n}{\ln \ln n}\geq 2$ for all $n> e$, we have:
\begin{align*}
&\Pr[X\geq (1+\delta)\cdot T^*] \leq e^{-T^*(1+\delta)\ln(1+\delta)/4}  \tag{\cref{def:chernoff}} \\
\implies &\Pr[X\geq \frac{10\ln n}{\ln \ln n}\cdot T^*] \leq e^{-T^*2\ln n} \tag{$\frac{10\ln n}{\ln\ln n}\ln\left( \frac{10 \ln n}{\ln \ln n} \right)\geq 8\ln n$} \\
\implies & \Pr[X\geq \frac{10\ln n}{\ln \ln n}\cdot T^*] \leq e^{-2\ln n} = \frac{1}{n^2} \tag{$T^*\geq 1$}
\end{align*}
Thus, \cref{lem:general:ratio} holds since $X=\abs{N_G(u)\cap S}$ for all $u\in L$.
\end{proof}

\paragraph{Wrapping Up.} Now, we are ready to prove \cref{thm:general}.

\begin{proof}[Proof of \cref{thm:general}]
By \cref{obs:pipage:running-time}, \cref{alg:pipage} returns a feasible integral solution in $\poly(n)$ times, where $n$ is the number of vertices in the input bipartite graph.
Recall that $S\subseteq R$ is the set of vertices returned by \cref{alg:pipage}.
By \cref{lem:general:ratio}, the vertex set $S$ has the following property:
$$
\Pr\left[ \abs{N_G(u)\cap S}\geq \frac{10 \ln n}{\ln \ln n}\cdot \opt \right] \leq \Pr\left[ \abs{N_G(u)\cap S} \geq \frac{10\ln n}{\ln \ln n}\cdot T^* \right] \leq \frac{1}{n^2}, \text{ for all } u\in L.
$$
The first inequality is due to the fact that $T^*$ is a lower bound of the optimal solution.
Then, by union bound, we have:
$$
\Pr\left[ \max_{u\in L}\abs{N_G(u)\cap S}\geq \frac{10 \ln n}{\ln \ln n}\cdot \opt \right] \leq \frac{1}{n}.
$$
Hence, the probability that the event $\max_{u\in L}\abs{N_G(u)\cap S}\leq \frac{10 \ln n}{\ln \ln n}\cdot \opt$ occurs is at least $1-\frac{1}{n}$.
Therefore, \cref{alg:pipage} is a $O(\frac{\log n}{\log \log n})$-approximate algorithm with running time $\poly(n)$.
\end{proof}

\subsection{Integrality Gap}
\label{sec:integrality-gap}

In this section, we show that our analysis is tight up to some constant factor by giving a hard instance to \eqref{Feas-LP}.
Formally, we mainly show \cref{thm:gap}.

\begin{theorem}
The natural linear programming formulation \eqref{Feas-LP} has an integrality gap of $\Omega(\frac{\log n}{\log \log n})$, where $n$ is the number of vertices in the input bipartite graph. 
\label{thm:gap}
\end{theorem}

\begin{figure}[htb]
    \centering
    \includegraphics[width=14.5cm]{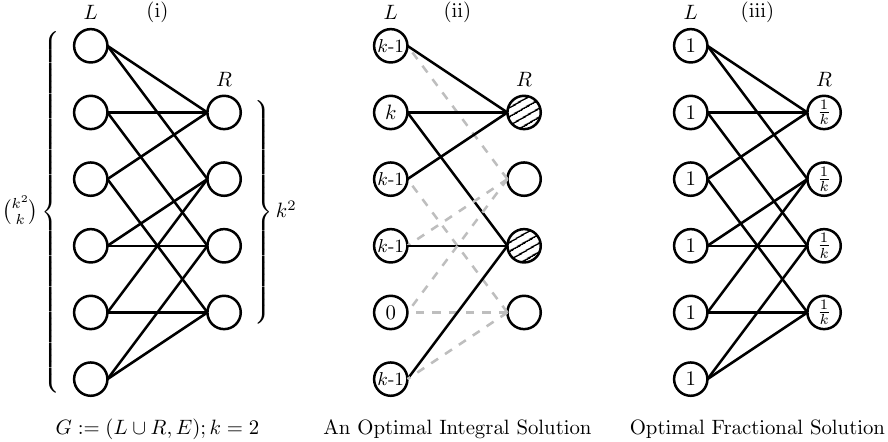}
    \caption{The integrality gap instance to \eqref{Feas-LP}. The figure shows an example with the demand $k=2$. Hence, there are $4$ vertices in $R$ and $\binom{4}{2}=6$ vertices in $L$; see subfigure (\rom{1}). The subfigure (\rom{2}) shows an optimal integral solution. By our construction, any vertex set of $R$ with cardinality $2$ is an optimal integral solution. In the example, we pick the first and third vertex of $R$, denoted by $S$. The solid lines of subfigure (\rom{2}) show the vertex-induced graph $G[S]$. The disagreement of all vertices in $L$ is marked inside of each vertex. The subfigure (\rom{3}) shows an optimal fractional solution in which the demand $k$ is equally distributed to each vertex in $R$. Thus, the disagreement of all vertices in $L$ is equal to each other, which is equal to $1$.}
    \label{fig:gap_instance}
\end{figure}

We first pick an integer $k$ as the demand, and then we construct a bipartite graph $G:=(L\cup R, E)$ such that $\abs{R}=k^2$ and $\abs{L}=\binom{k^2}{k}$.
Every vertex in $L$ stands for a possible selection of $k$ vertices from $R$.
Formally, consider an arbitrary subset of vertices $S\subseteq R$ with $\abs{S}=k$, there is a vertex $u$ in $L$.
For each vertex $v$ in $S$, there is an edge between $u$ and $v$.
See \cref{fig:gap_instance} for an example when $k=2$.
It is not hard to see that (\rom{1}) every vertex in $L$ has a degree of $k$; (2) every vertex in $R$ has a degree of $\binom{(k-1)^2}{k-1}$.

Before presenting the proof, we first discuss the relationship of parameters $n,k,\Delta$ in the constructed bipartite graph.
Recall that $n$ is the number of vertices in $L$.
Note $\binom{n}{m}\sim \frac{n^m}{m!}$ and $m!\sim \sqrt{2\pi m}(\frac{m}{e})^m$ (Stirling's approximation).
Thus, we have $\binom{k^2}{k} \sim \frac{1}{\sqrt{2\pi}}(ek)^{k-\frac{1}{2}}$.
Hence, we have $n=\abs{L\cup R} \approx \frac{1}{\sqrt{2\pi}}(ek)^{k-\frac{1}{2}}$.
Therefore, we have $k=\Theta(\frac{\log n}{\log \log n})$.
Note that $\binom{k^2}{k}\sim e^2\cdot k \cdot \binom{(k-1)^2}{k-1}$.
Therefore, the demand $k$ has the same relationship with the maximum degree $\Delta$, i.e., $k=\Theta(\frac{\log \Delta}{\log \log \Delta})$.
This bound shall be used in \cref{sec:degree=d} as a lower bound of our algorithm.

\begin{proof}[Proof of \cref{thm:gap}]
The proof is simple and based on the following two claims: (\rom{1}) any feasible integral solution has a maximum agreement of $k$; (\rom{2}) the fractional solution that equally distributes $k$ demands among all vertices in $R$ has a maximum disagreement of $1$, i.e., set $x_v=\frac{1}{k}$ for all $v\in R$.
Such an assignment is a feasible fractional solution since (\rom{1}) the solution has a maximum disagreement $1$, recall that we guess $T$ from $1$, so guessing cannot drop this value; (\rom{2}) $\sum_{v\in R}x_v=k^2\cdot\frac{1}{k}=k$.
Thus, the integrality gap is $k$, which is equal to $\Theta(\frac{\log n}{\log \log n})$.
\end{proof}

\section{Graphs with Bounded Degree}
\label{sec:degree=d}

In this section, we consider \FKSS/ on the graphs with a maximum degree $\Delta$ and mainly show that there is a $O(\log \Delta)$-approximate algorithm (\cref{thm:degree=d}).
Similar as \cref{sec:general}, to have a better understanding of our algorithmic idea, we shall focus on the case where all vertices in $R$ have the same weight.
We will extend our algorithm to the general \FKSS/ instances in \cref{sec:weight}.
We remark that our analysis is almost tight since the natural linear programming \eqref{Feas-LP} also has an integrality graph $\Omega(\frac{\log \Delta}{\log \log \Delta})$; see \cref{sec:integrality-gap}.

\begin{theorem}
Given any instance of \FKSS/ on graphs with a maximum degree $\Delta$, there is a randomized algorithm with running time $\poly(n)$ in the expectation that returns a $O(\log \Delta)$-approximate solution, where $n$ is the number of vertices in the input bipartite graph.
\label{thm:degree=d}
\end{theorem}

In practice, the maximum degree of many graphs is small compared to the number of vertices in the whole graph.
The graph with a maximum degree admits a trivial $\Delta$-approximate algorithm.
Namely, arbitrarily picking $k$-vertices from $R$ is a $\Delta$-approximate algorithm by observing the following two simple facts: (\rom{1}) the optimal solution to the given instance is at least $1$; (\rom{2}) the maximum disagreement of any algorithm is at most $\Delta$ since the maximum degree is $\Delta$.
However, if we use the algorithm for general graphs, we can only obtain a $O(\frac{\log n}{\log \log n})$-approximate solution; this is not even a constant factor when $\Delta$ is a constant.
In this section, we give a $O(\log \Delta)$-approximate algorithm, which significantly improves the ratio of the trivial $\Delta$-approximate algorithm.

\paragraph{Main Obstacles and Our Ideas.}
Our algorithm is still based on \eqref{Feas-LP}.
To get rid of the dependency of $n$ on the approximation ratio, there are three main obstacles that we need to overcome.
The first obstacle comes from the union bound we used in the proof of \cref{thm:general}.
For each vertex $u\in L$, we define the following event as a ``bad event'': the disagreement of vertex $u$ is larger than $\alpha\cdot\opt$ for some parameter $\alpha$. 
To ensure that these bad events do not occur at the same time, we need to use the union bound over all vertices in $L$; thus, we have to lose the factor $n$ on the approximation ratio.
However, if we do the analysis more carefully, we may realize that these bad events are not so independent.
For example, assuming that two vertices $u,v$ in $L$ have the same set of neighbors in $R$.
If the bad event of $u$ does not occur, this shall imply that the bad event of $v$ does not occur either.
Based on this observation, we may not need to do union bound over all vertices, so bypassing the dependency of $n$.
To this end, we shall use the powerful {\em Lovász Local Lemma}; see \cref{sec:LLL} for the definition of the lemma.

The next two obstacles come from the use of the Lovász Local Lemma.
The second obstacle is that the Lovász Local Lemma requires us to do independent rounding, but the independent rounding cannot ensure the feasibility of the solution, i.e., it cannot ensure the algorithm selects at least $k$ vertices.
Our idea to overcome this obstacle is that, aside from defining a bad event for the approximation ratio, we also define a bad event for the feasibility.
In this way, as long as the Lovász Local Lemma ensures all bad events do not occur, we obtain a solution that achieves the desired approximation ratio and satisfies the demand requirement simultaneously.
The third obstacle comes from the implementation of the above feasibility idea.
In other words, we need to find an appropriate definition for the feasibility bad events so that (\rom{1}) all feasibility bad events do not occur, ensuring a solution that satisfies the demand requirement; (\rom{2}) the defined bad events should not dependent on many other bad events to get rid of the dependency on $n$.
Our idea for this obstacle is the grouping technique; see \cref{def:badevents} for more details.

\subsection{Lovász Local Lemma}
\label{sec:LLL}

There are several versions of Lovász Local Lemma.
In our work, we shall use the {\em variable version} proposed by~\cite{DBLP:journals/jacm/MoserT10}.
In this version, there is an underlying family of mutually independent random variables on a common probability space, denoted by $\cX:=\set{X_1,\ldots,X_m}$.
Let $\cA:=\set{A_1,\ldots,A_n}$ be a set of bad events.
Each bad event $A_i$ is determined by a subset $\cA_i\subseteq \cX$ of variables in $\cX$.
The {\em dependency graph} $G:=(V,E)$ for $\cA$ is an undirected graph such that (\rom{1}) each vertex $v\in V$ corresponds a bad event $A_v$ in $\cA$; (\rom{2}) there is an edge between $u\in V$ and $v\in V$ if and only if $\cA_u\cap \cA_v \ne \emptyset$.
For each bad event $A_i$, let $\cN(A_i)\subseteq \cA$ be a set of bad events that are adjacent to $A_i$ in the dependency graph $G$.

\begin{lemma}[Lovász Local Lemma~\cite{DBLP:journals/jacm/MoserT10}]
If there exists a real number $x_i\in(0,1)$ for each bad event $A_i\in\cA$ such that $\Pr[A_i]\leq x_i \prod_{A_j\in \cN(A_i)}(1-x_j)$ for all $A_i\in\cA$, then (\rom{1}) there exists an assignment of the random variables in $\cX$ such that $\Pr[\wedge_{i\in[n]}\neg A_i]>0$; (\rom{2}) there exists an algorithm (\cref{alg:local-search}) that finds such an assignment for $\cX$ in expected time at most $\sum_{i\in[n]}\frac{x_i}{1-x_i}$. 
\label{lem:gLLL}
\end{lemma}

\begin{algorithm}[htb] 
\caption{~\cite{DBLP:journals/jacm/MoserT10} \texttt{LocalSearch($\cdot,\cdot$)}}
\label{alg:local-search}
\begin{algorithmic}[1]
\Require Random variable set $\cX=\set{X_1,\ldots,X_m}$; Bad event set $\cA:=\set{A_1,\ldots,A_n}$.
\Ensure An assignment to each random variable in $\cX$.
\State Pick an independent random assignment to each random variable in $\cX$.
\While{there exists a bad event occurs}
\State Pick an occurred bad event $A_j$ with the lowest index.
\State Resample all values of random variables in $\cA_j$.
\EndWhile
\State \Return The assignment $\fX$ of each variable in $\cX$.
\end{algorithmic}
\end{algorithm} 

Let $d$ be the maximum degree of $G$, i.e., $d:=\max_{i\in[n]}\abs{\cN(A_i)}$.
By setting $x_i=\frac{1}{d+1}$ for each $i\in [n]$, the Lovász Local Lemma can be simplified, which produces the {\em Symmetric Lovász Local Lemma}.
The symmetric Lovász Local Lemma was first proposed by~\cite{erdos1975problems}.

\begin{lemma}[Symmetric Lovász Local Lemma~\cite{erdos1975problems}]
Let $\cA=\set{A_1,\ldots,A_m}$ be a set of bad events with $\Pr[A_i]\leq p$ for all $i\in[n]$. 
If $e\cdot p\cdot (d+1) \leq 1$, then $\Pr[\wedge_{i\in[n]}\neg A_i]>0$.
\label{lem:sLLL}
\end{lemma}

To see that \cref{lem:gLLL} implies \cref{lem:sLLL}, one only needs to show $x_i \prod_{A_j\in \cN(A_i)}(1-x_j) \geq p$ when $x_i=\frac{1}{d+1}$ for all $i\in[n]$ since, in this way, we found the value for each $x_i$ that makes the condition of \cref{lem:gLLL} hold, i.e., $\Pr[A_i]\leq p \leq x_i \prod_{A_j\in \cN(A_i)}(1-x_j)$ for all $i\in[n]$.
The above inequality is true because of $ep(d+1)\leq 1$, i.e., for each $i\in[n]$,
$$
x_i\prod_{A_j\in\cN(A_i)}(1-x_j)\geq \frac{1}{d+1} \left( 1-\frac{1}{d+1} \right)^d > \frac{1}{(d+1)e}\geq p.
$$

\subsection{Rounding Algorithm}

\cref{alg:degree=d} is an independent rounding algorithm that is built based on the local search algorithm for Lovász Local Lemma (\cref{alg:local-search}).
As is typical, we interpret each variable in the optimal fractional solution $\bx^*=(x^*_v)_{v\in R}$ as the probability.
For each $v\in R$, we raise the probability of the event that selects $v$ to $O(\ln\Delta)\cdot x^*_v$ (see line~\ref{line:pv} of \cref{alg:degree=d}).
\cref{alg:degree=d} consists of three phases: (Phase 1) Fixed Choice Phase (lines~\ref{line:phase1:start}-\ref{line:phase1:end} of \cref{alg:degree=d}); (Phase 2) Local Search Phase (lines~\ref{line:phase2:start}-\ref{line:phase2:end} of \cref{alg:degree=d}); (Phase 3) One Vertex Phase (lines~\ref{line:phase3:start}-\ref{line:phase3:end} of \cref{alg:degree=d}).
In the first phase, we select all vertices whose probability is $1$.
If the number of selected vertices already satisfies the demand requirement (line~\ref{line:decision-phase2} of \cref{alg:degree=d}), then we stop; otherwise, we enter the second phase to select more vertices.
Due to some technical reason, \cref{alg:degree=d} may not be able to select sufficient vertices at the end of the second phase.
Fortunately, such a demand gap is at most $1$.
Thus, \cref{alg:degree=d} enters the third phase and arbitrarily selects one vertex.

The choice of the scaling factor $O(\ln\Delta)$ is crucial to the Lovász Local Lemma, i.e., it ensures that the probability of the bad events (which will be defined later in \cref{def:badevents}) satisfies the conditions stated in \cref{lem:sLLL}.
In this way, we can employ the local search algorithm (\cref{alg:local-search}) to find the right selection of vertices in $R$ that ensures all bad events do not occur.
Thus, we find the desired solution such that (\rom{1}) the selected vertices satisfy the demand requirement; (\rom{2}) the maximum disagreement of the solution is $O(\log \Delta)$-approximate.

\begin{algorithm}[htb] 
\caption{Rounding Algorithm for Degree Bounded Graphs}
\label{alg:degree=d}
\begin{algorithmic}[1]
\Require The bipartite graph $G:=(L\cup R, E)$; The maximum degree $\Delta$ of $G$; The demand $k$; The optimal fractional solution $\bx^*=(x^*_v)_{v\in R}$.
\Ensure A vertex set $S\subseteq R$ with $\abs{S}\geq k$.
\For{each vertex $v\in R$}
\State $p_v\leftarrow \min\{ 1, (x^*_v+\frac{1}{\Delta})\cdot 4 \cdot \ln (2e\Delta^2) \}$.
\label{line:pv}
\EndFor
\State $S_1 \leftarrow \set{v\in R\mid p_v = 1}$;
\label{line:phase1:start}
\Comment{Fixed Choice Phase}
\If {$\abs{S_1}\geq k$}
\label{line:decision-phase2}
\State \Return $S\leftarrow S_1$.
\EndIf
\label{line:phase1:end}
\If{$\abs{S_1}< k$}
\label{line:phase2:start}
\For{each vertex $v\in R \setminus S_1$}
\Comment{Local Search Phase}
\State Define $X_v\in\set{0,1}$ as a random variable such that $\Pr[X_v=1]=p_v$.
\EndFor
\State $\cX\leftarrow \set{X_v}_{v\in R\setminus S_1}$.
\State $\fX \leftarrow \texttt{LocalSearch}(\cX,\cA\cup\cB)$.
\Comment{Call \cref{alg:local-search}; see \cref{def:badevents} for $\cA$ and $\cB$}
\label{line:degree=d:badevents}
\State $S_2 \leftarrow \set{v\in R\setminus S_1 \mid X_v=1\text{ in }\fX}$.
\EndIf
\If{$\abs{S_1}+\abs{S_2}\geq k$}
\State \Return $S\leftarrow S_1 \cup S_2$.
\EndIf
\label{line:phase2:end}
\If{$\abs{S_1}+\abs{S_2}< k$}
\label{line:phase3:start}
\Comment{One Vertex Phase}
\State Pick an arbitrary vertex $v^*$ from $R\setminus (S_1\cup S_2)$.
\State \Return $S\leftarrow S_1 \cup S_2 \cup \set{v^*}$.
\EndIf
\label{line:phase3:end}
\end{algorithmic}
\end{algorithm} 

\subsection{Ratio Analysis}

We start by introducing some notations for the purpose of analysis.
Let $R \supseteq S:=S_1\cup S_2 $ be the set of vertices returned by \cref{alg:degree=d}, where $S_1$ and $S_2$ are the set of vertices selected in the first and second phase, respectively.
Let $\abs{S_1}:=k_1$ and $\abs{S_2}:=k_2$.
Recall that $T^*$ is the optimal fractional solution that is achieved by the solution $\bx^*=(x^*_v)_{v\in R}$.
Let $\opt$ be the maximum disagreement of the optimal solution.
Let $\alg$ be the maximum disagreement of the solution returned by \cref{alg:degree=d}.

\paragraph{Analysis Framework.}
To analyze the ratio, we distinguish three cases according to the existence of the second and third phases, i.e., Case (\rom{1}): \cref{alg:degree=d} ends at the first phase (\cref{lem:degree=d:phase1}); Case (\rom{2}): \cref{alg:degree=d} ends at the second phase (\cref{lem:degree=d:phase2}); Case (\rom{3}): \cref{alg:degree=d} ends at the third phase (\cref{lem:degree=d:phase3}).
The first phase of \cref{alg:degree=d} is equivalent to a simple deterministic rounding algorithm, and the third phase just picks at most one vertex; thus, their analysis is easy.
The analysis of the second phase depends on the Lovász Local Lemma.
To this end, we shall first define the bad events and show that the defined bad events achieve the desired approximation ratio (\cref{lem:degree=d:prop:S_2}) as well as (almost) satisfy the demand requirement (\cref{lem:degree=d:phase2:demand}).
Then, we will upper bound the probability that a bad event occurs (\cref{lem:degree=d:prob:B}, \ref{lem:degree=d:prob:A}) and show that \cref{alg:degree=d} finds an appropriate solution; this is the place where we use the Lovász Local Lemma.

We start with the easy case where \cref{alg:degree=d} does not enter the second phase; see \cref{lem:degree=d:phase1}.
We show that at the end of the second phase, the demand gap is at most $1$, and in the last phase, we pick an arbitrary vertex.
This will increase the value of \cref{alg:degree=d}'s solution by at most $1$.

\begin{lemma}
$\max_{u\in L}\dis_u(S_1) \leq 4\ln(2e\Delta^2)(\opt+1)$.
\label{lem:degree=d:prop:S_1}
\end{lemma}

\begin{proof}
The proof is standard and we state for completeness.
By definition, we have $\dis_u(S_1)=\abs{S_1\cap N_G(u)}$.
Observe that the first phase of \cref{alg:degree=d} is actually equivalent to the following simple deterministic rounding algorithm: define $\theta:=\frac{1}{4\ln(2e\Delta^2)}-\frac{1}{\Delta}$ as a threshold, then round all $x^*_v\geq \theta$ to $1$ and round all $x^*_v<\theta$ to $0$. 
Thus, $S_1$ is a set of vertices whose value is larger than $\theta$, i.e., $S_1=\set{v\in R\mid x^*_v \geq \theta}$.
Recall that for any $u\in L$, $\abs{N_G(u) \cap S_1}$ is the disagreement of vertex $u$, i.e., $\dis_u(S)=\abs{N_G(u) \cap S_1}$.
To show \cref{lem:degree=d:prop:S_1}, we need to upper bound the value of $\abs{N_G(u) \cap S_1}$ for all $u\in L$.
Thus, we consider any $u\in L$, we have:
$$
\opt\stackrel{(\rom{1})}{\geq} T^* \geq \sum_{v\in N_G(u)}x^*_v \stackrel{(\rom{2})}{\geq} \theta \cdot \abs{S_1\cap N_G(u)} = \frac{\abs{S_1\cap N_G(u)}}{4\ln(2e\Delta^2)}-\frac{\abs{S_1\cap N_G(u)}}{\Delta} \stackrel{(\rom{3})}{\geq} \frac{\dis_u(S_1)}{4\ln(2e\Delta^2)}-\frac{\Delta}{\Delta}.
$$
The inequality (\rom{1}) above is due to the fact that $T^*$ is a lower bound of the optimal integral solution.
The inequality (\rom{2}) is due to the fact that, for all vertices in $v\in S_1\cap N_G(u)$, $x^*_v\geq \theta$.
The inequality (\rom{3}) is due to the fact that $\abs{S_1\cap N_G(u)}\leq \abs{N_G(u)} \leq \Delta$.
Therefore, we have $\dis_u(S_1)\leq 4\ln(2e\Delta^2)(\opt+1)$ for all $u\in L$.
Thus, \cref{lem:degree=d:prop:S_1} holds.    
\end{proof}

\begin{lemma}
If \cref{alg:degree=d} does not enter the second phase, then the following two claims are true: (\rom{1}) $S=S_1$ is a feasible solution; (\rom{2}) $\alg\leq 4\cdot\ln(2e\Delta^2)\cdot (\opt+1)$.  
\label{lem:degree=d:phase1}
\end{lemma}

\begin{proof}
If \cref{alg:degree=d} does not enter the second phase, then $\abs{S_1}\geq k$ by line~\ref{line:decision-phase2} of \cref{alg:degree=d}, i.e., the size of $S_1$ satisfies the demand requirement.
Thus, the first claim of \cref{lem:degree=d:phase1} holds.
The second claim directly follows from \cref{lem:degree=d:prop:S_1}.
Since \cref{alg:degree=d} does not enter the second phase, then $S_1$ is the solution returned by the algorithm.
Thus, we have $\alg=\max_{u\in L}\dis_u(S_1)\leq 4\ln(2e\Delta^2)(\opt+1)$, which proves the second claim of \cref{lem:degree=d:phase1}.
\end{proof}

Now, we focus on the second case where \cref{alg:degree=d} has the second phase, which implies that $k_1< k$.
Let $R':=R\setminus S_1$, and if a vertex in $L$ has no neighbors in $R'$, we delete such a vertex.
Let $L'$ be the set of remaining vertices in $L$.
We shall focus on the subgraph $G':=(L'\cup R', E')$.
Note that for each $v\in R'$, we have $(x^*_v+\frac{1}{\Delta})\cdot 4\ln(2e\Delta^2)<1$.
For each vertex $v\in R'$, define a random variable $X_v\in\set{0,1}$ to indicate whether the vertex $v$ is added to $S_2$.
Let $\cX:=\set{X_v}_{v\in R'}$.
We now give the definition of the {\em bad events} based on the random variable set $\cX$, which is crucial to the Lovász Local Lemma.

\begin{figure}[htb]
    \centering
    \includegraphics[width=15cm]{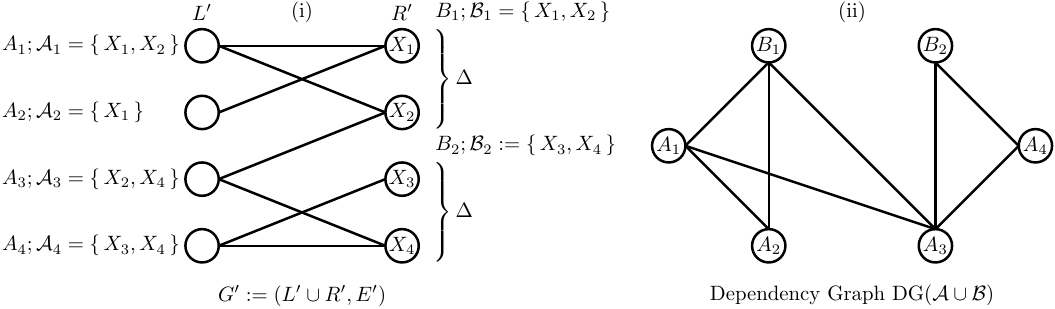}
    \caption{An example for the bad events (\cref{def:badevents}) and their dependency graph (\cref{def:dependency-graph}). The remaining graph $G'$ has four vertices in $L'$ and four vertices in $R'$. Assuming that the maximum degree $\Delta$ of the original graph is $2$. The bad events are shown in the subfigure (\rom{1}). For each vertex $v$ in $R'$, there is a random variable $X_v\in\set{0,1}$ indicating whether $v$ is added to the solution. All bad events in $\cA\cup\cB$ are defined over $\cX:=\set{X_v}_{v\in R}$. For each vertex $u\in L'$, we have one bad event $A_u$ in $\cA$, and $A_u$ is determined by a set $\cA_u$ of random variables in $\cX$. For example, bad event $A_1$ is determined by $\set{X_1,X_2}$ which are neighbors of vertex $1$. We group all vertices in $R'$ into several subgroups, each of which consists of at most $\Delta$ vertices in $R'$. Then, we shall first define a bad event for each subgroup whose size is exactly $\Delta$. For the last subgroup, whose size may be smaller than $\Delta$, we might have a bad event depending on some condition. The dependency graph of $\cA\cup\cB$ is shown in the subfigure (\rom{2}). The dependency graph describes the relationship between all bad events. For example, the occurrence of bad event $B_1$ depends on the occurrence of bad events $A_1,A_2,A_3$.}
    \label{fig:bad-events}
\end{figure}

\paragraph{Bad Events.}
Intuitively, the definition of bad events needs to ensure that when all of them do not occur, the following two statements are true: (\rom{1}) the selected vertices in the second phase satisfy the remaining demands; (\rom{2}) the maximum disagreement of the selected vertices in the second phase is $O(\log \Delta)\cdot \opt$.
This motivates us to define two groups of bad events: {\em performance bad events} $\cA$ and {\em feasibility bad events} $\cB$.
The bad event in $\cA$ is used to ensure that the selected vertices are able to achieve a good ratio.
To this end, for each vertex in $u\in L'$, we define a bad event $A_u$ stands for $\abs{N_{G'}(u)\cap S_2}> O(\log\Delta)\cdot\opt$, where $N_{G'}(u)$ is a set of neighbors of vertex $u$ in graph $G'$.
The bad event in $\cB$ is used to ensure that the number of vertices in $S_2$ satisfies the remaining demand requirement, i.e., $\abs{S_2}\geq k-k_1$.
To this end, we group vertices in $R'$ into several subgroups according to the index of vertices such that each subgraph consists of $\Delta$ vertices from $R'$.
Let $D_1,\ldots,D_{\ell}\subseteq R'$ be these subgroups. 
Note that the size of the last subgroup $D_{\ell}$ may be smaller than $\Delta$, i.e., $\abs{D_{\ell}}<\Delta$.
For each subgroup $D_i,i\in[\ell-1]$, we define a bad event $B_i$ stands for $\abs{D_i\cap S_2} <\sum_{u\in D_i}x_u^*$; its meaning will be clear in the proof of \cref{lem:degree=d:phase2:demand}.
Intuitively, this inequality requires that the number of selected vertices in $D_i$ is at least $\sum_{u\in D_i}x_u^*$.
For the last subgroup $D_{\ell}$, we shall not define a bad event if $\sum_{u\in D_{\ell}}(x_u^*+\frac{1}{\Delta})<1$; otherwise, $D_{\ell}$ also has a bad event $B_{\ell}$; its meaning will be clear in the proof of \cref{lem:degree=d:prob:B}.
Intuitively, if $D_{\ell}$ has no bad event, then $\sum_{u\in D_{\ell}}x^*_u$ is small; thus, we will not create a big demand gap in the case where no vertex is selected from $D_{\ell}$.
See \cref{def:badevents} for the formal definition of the bad events.
An example can be found in \cref{fig:bad-events}.

\begin{definition}[Bad Events]
The bad event set $\cA\cup\cB$ consists of two types of bad events, where bad events $\cA$ and $\cB$ are called performance bad events (P-BE) and feasibility bad events (F-BE), respectively.
\begin{enumerate}[label=(S-SS*),leftmargin=*,align=left]
    \item[(P-BE)] For each vertex $u\in L'$, we have a bad event $A_u$ in $\cA$. Define $\cA_u\subseteq \cX$ as a subset of random variables, each of which corresponds to a vertex in $N_{G'}(u)$. We say $A_u$ occurs if:
    $$
    \sum_{i\in\cA_u} X_i \geq 8 \cdot \ln(2e\Delta^2) \cdot (T^*+1).
    $$
    \item[(F-BE)] For each subgroup $D_j,j\in[\ell-1]$, we have a bad event in $\cB$. The last subgraph $D_{\ell}$ has a bad event if $\sum_{u\in D_{\ell}}(x^*_u+\frac{1}{\Delta})<1$; otherwise, $D_{\ell}$ does not have a bad event. For each bad event $B_j\in\cB$, define $\cB_j\subseteq \cX$ as a subset of random variables, each of which corresponds to a vertex in $D_j$. We say $B_j$ occurs if:
    $$
    \sum_{i\in\cB_j} X_i < \sum_{u\in D_j}x^*_u.
    $$
\end{enumerate}
\label{def:badevents}
\end{definition}

\begin{definition}[Dependency Graph]
The dependency graph of all bad events in $\cA\cup\cB$, denoted by $\dg(\cA\cup\cB)$ is a undirected graph where: (\rom{1}) each vertex in $\dg(\cA\cup\cB)$ corresponds to a bad event in $\cA\cup\cB$; (\rom{2}) consider two vertices $i$ and $j$, let $C_i,C_j\in\cA\cup\cB$ be the corresponding bad event. There is an edge between $i,j$ if and only if $\cC_i\cap\cC_j\ne\emptyset$, where $\cC_i,\cC_j\subseteq\cX$ are the corresponding random variable sets.
\label{def:dependency-graph}
\end{definition}

\cref{lem:degree=d:dependency-graph} gives a lower and upper bound of the maximum degree of the dependency graph.
These two bounds shall be used later to connect our problem and Lovász Local Lemma.

\begin{lemma}
Let $d$ be the maximum degree of the dependency graph $\dg(\cA\cup\cB)$.
Then, the following two inequalities hold: (\rom{1}) $d\geq 1$; (\rom{2}) $d\leq \Delta^2$, where $\Delta$ is the maximum degree of the input bipartite graph.
\label{lem:degree=d:dependency-graph}
\end{lemma}

\begin{proof}
We first lower bound the value of $d$.
Observe that if $\dg(\cA\cup\cB)$ contains no isolated vertex, then $d\geq 1$.
Consider an arbitrary vertex $u\in L'$; if there is an edge between $u$ and some vertex $v\in R'$, then the degree of the corresponding bad event $A_u$ is at least $1$.
On the other hand, consider any vertex $v\in R'$ and assume that $v\in D_{j}$; if there is an edge between $v$ and some vertex $u\in L'$, the degree of $D_j$'s bad event is at least $1$.
Hence, $\dg(\cA\cup\cB)$ has no isolated vertex as long as there is no isolated vertex in $G':=(L'\cup R',E')$.
$G'$ has no isolated vertex because we already remove all isolated vertices from $G$.
Therefore, the first claim of \cref{lem:degree=d:dependency-graph} holds.

We now upper bound the value of $d$.
Consider an arbitrary vertex $u\in L'$ and its corresponding bad event $A_u$.
Note that $\cA_u$ consists of at most $\Delta$ random variables from $\cX$, each of which can be contained in at most $\Delta-1$ different bad events in $\cA$ and at most one bad event in $\cB$.
Thus, the degree of any bad event in $\cA$ is at most $\Delta^2$.
For any bad event $B_i\in\cB$, $\cB_i$ consists of at most $\Delta$ random variables from $\cX$, each of which can be contained in at most $\Delta$ bad events in $\cA$ and $0$ bad event in $\cB$.
Thus, the degree of any bad event in $\cB$ is at most $\Delta^2$.
Hence, the second claim of \cref{lem:degree=d:dependency-graph} holds.
\end{proof}

In the following, we shall split the proof into two parts.
In the first part (Part \Rom{1}), we focus on an assignment $\fX$ of random variables in $\cX$ that makes all bad events in $\cA\cup\cB$ not occur.
We define $S_2:=\set{v\in R' \mid X_v=1 \text{ in }\fX}$.
We show that $S_2$ is a set of vertices that satisfies the claimed property.
In the second part (Part \Rom{2}), we prove that \cref{alg:degree=d} is able to return such an assignment in the desired running time.
The proofs in this part shall built on the Lovász Local Lemma stated in \cref{lem:gLLL} and \cref{lem:sLLL}.

\paragraph{Part \Rom{1}: Properties of $\fX$ and $S_2$.}
We start with the correctness of the feasibility bad events' definition.
Formally, we shall show that when all bad events in $\cB$ do not occur, the demand gap is at most $1$ at the end of the second phase.

\begin{lemma}
If \cref{alg:degree=d} enters the second phase and all bad events in $\cB$ do not occur, then $\abs{S_1}+\abs{S_2} \geq k-1$.
\label{lem:degree=d:phase2:demand}
\end{lemma}

\begin{proof}
By the definition of bad events in $\cB$, each subgroup in $\set{D_1,\ldots,D_{\ell-1}}$ corresponds to a bad event, and $D_{\ell}$ might have a bad event depending on the values of $x_u^*$ in $D_{\ell}$.
Hence, we distinguish two cases.

In the case where $D_{\ell}$ has a bad event in $\cB$, then every subgroup has a bad event in $\cB$.
Since all bad events do not occur, we have $\abs{D_i\cap S_2} \geq \sum_{u\in D_i}x_u^*$ for all $i\in[\ell]$.
Note that $D_i\cap D_j=\emptyset$ for any $i,j\in[\ell]$.
Thus, we have $\abs{S_2}\geq \sum_{u\in R'}x_u^*$.
Recall that $k_1$ is the size of $S_1$ and $k_1<k$ since \cref{alg:degree=d} enters the second phase.
Observe that if $\sum_{u\in R'}x_u^* \geq k-k_1$, then $\abs{S_2}\geq \sum_{u\in R'}x_u^*\geq k-k_1$.
Thus, \cref{lem:degree=d:phase2:demand} holds since $\abs{S_1}+\abs{S_2}\geq k$.
The inequality $\sum_{u\in R'}x_u^* \geq k-k_1$ holds since 
(\rom{1}) $\sum_{u\in R}x_u^*=k$; (\rom{2}) $R'=R\setminus S_1$; (\rom{3}) $\abs{S_1}=k_1$; (\rom{4}) $x_u^*\leq 1$ for all $u\in R$.
In other words, removing $S_1$ from $R$ reduces the value of $\sum_{u\in R}x_u^*$ at most $k_1$ (due to (\rom{3}) and (\rom{4})), so the remaining part $\sum_{u\in R'}x_u^*$ has a value of at least $k-k_1$ (due to (\rom{1}) and (\rom{2})).

In the case where $D_{\ell}$ does not have a bad event in $\cB$, then only the subgroups in $\set{D_1,\ldots,D_{\ell-1}}$ have a bad event in $\cB$.
Since all bad events do not occur, we have $\abs{D_i\cap S_2}\geq \sum_{u\in D_i}x_u^*$ for all $i\in[\ell-1]$.
Since $D_i\cap D_j=\emptyset$ for any $i,j\in[\ell]$, we have $\abs{S_2}\geq\sum_{u\in R'\setminus D_{\ell}}x_u^*$.
Observe that if $\sum_{u\in R'\setminus D_{\ell}}x_u^*\geq k-k_1-1$, then $\abs{S_2}\geq k-k_1-1$.
Thus, \cref{lem:degree=d:phase2:demand} holds since $\abs{S_1}+\abs{S_2}\geq k-1$.
Note that $D_{\ell}$ does not have a bad event in $\cB$.
Hence, $\sum_{u\in D_{\ell}}(x_u^*+\frac{1}{\Delta})<1$ by \cref{def:badevents}.
This implies that: (\rom{5}) $\sum_{u\in D_{\ell}}x^*_u<1$.
Similar as the first case, $\sum_{u\in R'\setminus D_{\ell}}x_u^*\geq k-k_1-1$ holds, because removing $S_1$ from $R$ reduces the value of $\sum_{u\in R}x^*_u$ at most $k_1$ (due to (\rom{3}) and (\rom{4})), further removing $D_{\ell}$ from $R'$ reduces the value of $\sum_{u\in R'}x^*_u$ at most $1$ (due to (\rom{5})), so the remaining part $\sum_{u\in R'\setminus D_{\ell}}x_u^*$ has a value of at least $k-k_1-1$ (due to (\rom{1}) and (\rom{2})).
\end{proof}

\begin{lemma}
$\max_{u\in L'}\abs{S_2\cap N_{G'}(u)}\leq 8\ln(2e\Delta^2)(\opt+1)$.
\label{lem:degree=d:prop:S_2}
\end{lemma}
\begin{proof}
By the definition of performance bad events (\cref{def:badevents}), if all bad events in $\cA$ do not occur, we have $\sum_{i\in\cA_u}X_i< 8\ln(2e\Delta^2)(T^*+1)$ for all $u\in L'$.
Note that $\sum_{i\in\cA_u}X_i=\abs{S_2\cap N_{G'}(u)}$ for all $u\in L'$ by the definition of $S_2$.
Thus, we have $\max_{u\in L'}\abs{S_2\cap N_{G'}(u)}\leq 8\ln(2e\Delta^2)(T^*+1)\leq 8\ln(2e\Delta^2)(\opt+1)$.   
\end{proof}

\begin{lemma}
$\max_{u\in L}\dis_u(S_1\cup S_2)\leq 12\ln(2e\Delta^2)(\opt+1)$.
\label{lem:degree=d:prop:S_1+S_2}
\end{lemma}

\begin{proof}
By definition, we have $\dis_u(S_1\cup S_2)=\abs{(S_1\cup S_2)\cap N_{G}(u)}$.
To show \cref{lem:degree=d:prop:S_1+S_2}, we upper bound the value of $\max_{u\in L\setminus L'}\abs{(S_1\cup S_2)\cap N_{G}(u)}$ and the value of $\max_{u\in L'}\abs{(S_1\cup S_2)\cap N_{G}(u)}$ separately.
Based on \cref{clm:degree=d:phase1+2:L-L'} and \cref{clm:degree=d:phase1+2:L'}, we have
\begin{align*}
\max_{u\in L}\abs{(S_1\cup S_2)\cap N_G(u)}
&=\max\{\max_{u\in L\setminus L'}\abs{(S_1\cup S_2)\cap N_G(u)},\max_{u\in L'}\abs{(S_1\cup S_2)\cap N_G(u)}\} \\
&\leq 12\ln(2e\Delta^2)(\opt+1). \tag{\cref{clm:degree=d:phase1+2:L-L'}, \ref{clm:degree=d:phase1+2:L'}}
\end{align*}
This proves the second claim of \cref{lem:degree=d:phase2}.

\begin{claim}
$\max_{u\in L\setminus L'}\abs{(S_1\cup S_2)\cap N_G(u)} \leq 4\ln(2e\Delta^2)(\opt+1)$.   
\label{clm:degree=d:phase1+2:L-L'}
\end{claim}

\begin{proof}
Recall that $L'$ is produced form $L$ by removing those vertices which have no neighbors in $R\setminus S_1$, i.e., $L\setminus L'=\set{v\in L \mid N_{G}(v)\subseteq S_1}$.   
Thus, the disagreement of all vertices in $L\setminus L'$ remains the same after the first phase ends even if \cref{alg:degree=d} selects more vertices in the second and third phases.
Hence, by \cref{lem:degree=d:prop:S_1}, we have:
$$
\max_{u\in L\setminus L'}\abs{(S_1\cup S_2)\cap N_G(u)}=\max_{u\in L\setminus L'}\abs{S_1\cap N_G(u)}\leq \max_{u\in L}\abs{S_1\cap N_G(u)}\leq 4\ln(2e\Delta^2)(\opt+1).
$$
\end{proof}

\begin{claim}
$\max_{u\in L'}\abs{(S_1\cup S_2)\cap N_G(u)}\leq 12\ln(2e\Delta^2)(\opt+1)$.
\label{clm:degree=d:phase1+2:L'}
\end{claim}
\begin{proof}
Consider any $u\in L'$, $N_{G'}(u)$ can be produced from $N_G(u)$ by removing all vertices in $S_1$.
Hence, $S_2\cap N_{G'}(u)$ is the same as the $S_2\cap N_{G}(u)$.
Hence, by \cref{lem:degree=d:prop:S_1} and \cref{lem:degree=d:prop:S_2}, we have:
\begin{align*}
\max_{u\in L'}\abs{(S_1\cup S_2)\cap N_G(u)}
&\leq \max_{u\in L'}\abs{S_1\cap N_{G}(u)}+\max_{u\in L'}\abs{S_2\cap N_{G}(u)} \\
&= \max_{u\in L'}\abs{S_1\cap N_{G}(u)}+\max_{u\in L'}\abs{S_2\cap N_{G'}(u)} \tag{$S_2\cap N_{G'}(u)=S_2\cap N_{G}(u)$} \\
&\leq 4\ln(2e\Delta^2)(\opt+1)+8\ln(2e\Delta^2)(\opt+1) \tag{\cref{lem:degree=d:prop:S_1}, \ref{lem:degree=d:prop:S_2}} \\
&=12\ln(2e\Delta^2)(\opt+1).
\end{align*}
\end{proof}    
\end{proof}

\begin{lemma}
If \cref{alg:degree=d} enters the second phase but does not enter the third phase, and $S_2$ is a solution that makes all bad events in $\cA\cup\cB$ do not occur, then the following two claims are true: (\rom{1}) $\abs{S_1\cup S_2}\geq k$; (\rom{2}) $\alg \leq 12\cdot\ln(2e\Delta^2)\cdot (\opt+1)$.
\label{lem:degree=d:phase2}
\end{lemma}

\begin{proof}
By the definition of \cref{alg:degree=d}, if the algorithm does not enter the third phase, then $\abs{S_1}+\abs{S_2}$ satisfies the demand requirement.
Thus, the first claim holds.
In this case, $S_1\cup S_2$ is the solution returned by \cref{alg:degree=d}.
Thus, we have $\alg=\max_{u\in L}\dis_u(S_1\cup S_2)\leq 12\ln(2e\Delta^2)(\opt+1)$ by \cref{lem:degree=d:prop:S_1+S_2}, which proves the second claim the lemma.
\end{proof}

\begin{lemma}
If \cref{alg:degree=d} enters the third phase and selects $v^*$ in the third phase.
Assume that $S_2$ is a solution that makes all bad events in $\cA\cup\cB$ do not occur, then the following two claims are true: (\rom{1}) $S_1\cup S_2 \cup\set{v^*}$ is a feasible solution; (\rom{2}) $\alg\leq 12\cdot\ln(2e\Delta^2)\cdot(\opt+2)$.
\label{lem:degree=d:phase3}
\end{lemma}
\begin{proof}
By \cref{lem:degree=d:phase2:demand}, we have $\abs{S_1}+\abs{S_2}\geq k-1$.
Since \cref{alg:degree=d} enters the third phase and selects $v^*$ from $R\setminus (S_1\cup S_2)$.
Thus, the solution $S$ returned by \cref{alg:degree=d} is $S_1\cup S_2\cup\set{v^*}$, whose size is larger than the demand requirement.
Hence, the first claim of \cref{lem:degree=d:phase3} holds.
By \cref{lem:degree=d:prop:S_1+S_2}, we have:
\begin{align*}
\max_{u\in L}\abs{(S_1\cup S_2\cup \set{v^*}) \cap N_{G}(u)}
&\leq \max_{u\in L}\abs{(S_1\cup S_2)\cap N_{G}(u)} + \max_{u\in L}\abs{\set{v^*}\cap N_G(u)} \\
&\leq 12\ln(2e\Delta^2)(\opt+1) + 1 \tag{\cref{lem:degree=d:prop:S_1+S_2}} \\
&\leq 12\ln(2e\Delta^2)(\opt+2) \tag{$12\ln(2e\Delta^2)\geq 1$}
\end{align*}

\end{proof}

\paragraph{Part \Rom{2}: Finding $\fX$ and $S_2$.}
In this part, we upper bound the probability that a bad event occurs (\cref{lem:degree=d:prob:B} and \cref{lem:degree=d:prob:A}).
This shall prove that our bad events satisfy the condition of the Lovász Local Lemma (\cref{lem:sLLL}).
And then, we use \cref{lem:gLLL} to bound the running time of \cref{alg:degree=d}.

\begin{lemma}
For all bad events $B_i\in\cB$, $\Pr[B_i\text{ occurs}] \leq \frac{1}{2 e d}$, where $d$ is the maximum degree of the dependency graph $\dg(\cA\cup\cB)$.
\label{lem:degree=d:prob:B}
\end{lemma}

\begin{proof}
We shall use the left tail Chernoff bound, which is stated in the following for completeness.
Let $X_1,\ldots,X_n$ be independent random variables such that $X_i\in[0,1]$ for all $i\in [n]$.
Define $X:=\sum_{i=1}^{n}X_i$ and let $\mu:=\E[X]$ be the expectation of $X$.
Let $\cmu$ be some lower bound of $\mu$, i.e., $\cmu\leq \mu$.
Then, for any $\delta\in(0,1)$, we have
\begin{equation}
\Pr[X\leq (1-\delta)\cdot\cmu] \leq e^{-\delta^2\cdot \mu/2}\leq e^{-\delta^2\cdot \cmu/2}.    
\label{equ:chernoff:left}
\end{equation}
In our problem, recall that for each vertex $v\in R'$, the random variable $X_v\in\set{0,1}$ is used to indicate whether the vertex $v$ is added to the solution $S_2$.
$\cX:=\set{X_v}_{v\in R'}$ is defined as a collection of all random variables above.
Note that $\Pr[X_v=1]=(x_v^*+\frac{1}{\Delta})\cdot 4 \cdot \ln(2e\Delta^2)<1$.
Consider an arbitrary bad event $B_i\in\cB$, recall that 
$D_i\subseteq R'$ is the corresponding vertex set and $\cB_i\subseteq \cX$ is the corresponding random variable set.
Define $X:=\sum_{v\in D_i}X_v$.
Note that 
\begin{equation}
\E[X]=4\cdot\ln(2e\Delta^2)\sum_{v\in D_i}\left(x^*_v+\frac{1}{\Delta}\right)= 4\cdot\ln(2e\Delta^2)\left(\frac{\abs{D_i}}{\Delta}+\sum_{v\in D_i}x^*_v\right)\geq 4\cdot \ln(2e\Delta^2).   
\label{equ:B:expectation}
\end{equation}
The last inequality\footnote{We remark that this is the reason why we need to add $\frac{1}{\Delta}$ to the probability and why we need to drop the last subgroup $D_{\ell}$ if $\sum_{v\in D_{\ell}}(x^*_v+\frac{1}{\Delta})<1$.} is due to $\abs{D_i}=\Delta$ for all $i\in[\ell-1]$.
For the last subgraph $D_{\ell}$, we have $\sum_{v\in D_{\ell}}(x_v^*+\frac{1}{\Delta})\geq 1$ since $D_{\ell}$ has a bad event in $\cB$ by \cref{def:badevents}.
Now, we use the Chernoff bound above by setting $\delta=\frac{3}{4}$.
\begin{align*}
&\Pr[X\leq (1-\delta)\cdot \cmu] \leq e^{-\delta^2\cdot \cmu/2} \tag{Chernoff Bound \cref{equ:chernoff:left}} \\ 
\implies & \Pr\left[X\leq \frac{1}{4} \cdot \cmu \right] \leq e^{-9\cdot \cmu/32} \tag{$\delta=\frac{3}{4}$} \\
\implies & \Pr\left[X\leq \ln(2e\Delta^2)\left(\sum_{v\in D_i}x_v^*+\frac{\abs{D_i}}{\Delta}\right)\right] \leq e^{-9\cdot\ln(2e\Delta^2)/8} \tag{\cref{equ:B:expectation}} \\
\implies & \Pr\left[X\leq \ln(2e\Delta^2)\left(\sum_{v\in D_i}x_v^*+\frac{\abs{D_i}}{\Delta}\right)\right] \leq \left(\frac{1}{e}\right)^{\ln(2e\Delta^2)} \\
\implies & \Pr\left[X\leq \ln(2e\Delta^2)\left(\sum_{v\in D_i}x_v^*+\frac{\abs{D_i}}{\Delta}\right)\right] \leq \frac{1}{2e\Delta^2} \leq \frac{1}{2ed} \tag{$d\leq\Delta^2$ in \cref{lem:degree=d:dependency-graph}} \\
\end{align*}

By \cref{def:badevents}, the bad event $B_i$ occurs means that $\abs{D_i\cap S_2}<\sum_{u\in D_i}x_u^*$, i.e., $X<\sum_{u\in D_i}x_u^*$.
Since $\ln(2e\Delta^2)\geq 1$, for any $B_i\in\cB$, we have 
$$
\Pr[B_i\text{ occurs}]=\Pr\left[X<\sum_{u\in D_i}x^*_u\right]\leq \Pr\left[X\leq \ln(2e\Delta^2)\left(\sum_{v\in D_i}x_v^*+\frac{\abs{D_i}}{\Delta}\right)\right] \leq \frac{1}{2ed}.
$$
\end{proof}

\begin{lemma}
For all bad events $A_i\in\cA$, $\Pr[A_i\text{ occurs}]\leq\frac{1}{2ed}$, where $d$ is the maximum degree of the dependency graph $\dg(\cA\cup\cB)$.    
\label{lem:degree=d:prob:A}
\end{lemma}

\begin{proof}
We shall use the right tail Chernoff bound, which is stated in the following for completeness.    
Let $X_1,\ldots,X_n$ be independent random variables such that $X_i\in[0,1]$ for all $i\in[n]$.
Define $X:=\sum_{i=1}^{n}X_i$ and let $\mu:=\E[X]$ be the expectation of $X$.
Let $\hmu$ be some upper bound of $\mu$, i.e., $\mu\leq\hmu$.
Then, for any $\delta\geq 1$, we have
\begin{equation}
\Pr[X\geq (1+\delta)\cdot\hmu] \leq e^{-(1+\delta) \cdot \ln(1+\delta)\cdot \hmu /4}.    
\label{equ:chernoff:right}
\end{equation}
In our problem, recall that for each vertex $v\in R'$, the random variable $X_v\in\set{0,1}$ is used to indicate whether the vertex $v$ is added to the solution $S_2$.
$\cX:=\set{X_v}_{v\in R'}$ is defined as a collection of all random variables above.
Note that $\Pr[X_v=1]=(x^*_v+\frac{1}{\Delta})\cdot 4 \cdot \ln(2e\Delta^2)<1$.
Note that each vertex $i\in L'$ corresponds to a bad event $A_i\in\cA$, recall that $N_{G'}(i)\subseteq R'$ be the set of neighbors of vertex $i$ in graph $G':=G[R']$ and $\cA_i\subseteq\cX$ is the corresponding random variable set.
We focus an arbitrary bad event $A_i$, define $X:=\sum_{v\in N_{G'}(i)}X_v$.
Note that 
\begin{align*}
\E[X]
&=4\cdot \ln(2e\Delta^2)\sum_{v\in N_{G'}(i)}(x^*_v+\frac{1}{\Delta})
=4\cdot \ln(2e\Delta^2)\cdot\left(\frac{\abs{N_{G'}(i)}}{\Delta}+\sum_{v\in N_{G'}(i)}x^*_v\right)  \\
&\leq 4\cdot \ln(2e\Delta^2)\cdot\left(1+\sum_{v\in N_{G'}(i)}x^*_v\right) \tag{$\abs{N_{G'}(i)}\leq \abs{N_{G}(i)}\leq\Delta$} \\
&\leq 4\cdot \ln(2e\Delta^2)\cdot(1+T^*) \tag{$\sum_{v\in N_{G'}(i)}x^*_v\leq \sum_{v\in N_{G}(i)}x^*_v \leq T^*$} \\
\end{align*}
Now, we use the Chernoff bound above by setting $\delta=1$.
\begin{align*}
&\Pr[X\geq (1+\delta)\cdot\hmu] \leq e^{-(1+\delta) \cdot \ln(1+\delta)\cdot \hmu /4} \tag{Chernoff Bound \cref{equ:chernoff:right}} \\
\implies & \Pr[X\geq 2\cdot\hmu] \leq e^{-2\ln2\cdot\hmu/4} \tag{$\delta=1$} \\
\implies &\Pr[X\geq 8\cdot\ln(2e\Delta^2)\cdot(1+T^*)] \leq e^{-2\cdot\ln2\cdot(\ln(2e\Delta^2))\cdot(1+T^*)} \\
\implies &\Pr[X\geq 8\cdot\ln(2e\Delta^2)\cdot(1+T^*)] \leq e^{-(\ln(2e\Delta^2))\cdot(1+T^*)} \tag{$2\ln2\geq 1$} \\
\implies &\Pr[X\geq 8\cdot\ln(2e\Delta^2)\cdot(1+T^*)] \leq e^{-\ln(2e\Delta^2)} = \left( \frac{1}{e} \right)^{\ln(2e\Delta^2)} \tag{$1+T^*\geq 1$} \\
\implies &\Pr[X\geq 8\cdot\ln(2e\Delta^2)\cdot(1+T^*)] \leq \frac{1}{2e\Delta^2} \leq \frac{1}{2ed} \tag{$\Delta^2\geq d$ in \cref{lem:degree=d:dependency-graph}}
\end{align*}
By \cref{def:badevents}, the bad event $A_i$ occurs means that $\abs{N_{G'}(i)\cap S_2} \geq 8\ln(2e\Delta^2)(T^*+1)$, i.e., $X\geq 8\ln(2e\Delta^2)(T^*+1)$.
Thus, for any $A_i\in\cA$, we have
$$
\Pr[A_i\text{ occurs}]=\Pr[X\geq 8\ln(2e\Delta^2)(T^*+1)]\leq\frac{1}{2ed}.
$$
\end{proof}

\paragraph{Wrapping Up.}
Now, we are ready to prove \cref{thm:degree=d}.
\begin{proof}[Proof of \cref{thm:degree=d}]
By \cref{lem:degree=d:phase1}, \cref{lem:degree=d:phase2}, and \cref{lem:degree=d:phase3}, we know that if \cref{alg:degree=d} returns a solution that makes all bad events not occur, then \cref{alg:degree=d} is a $12\ln(2e\Delta^2)(\opt+2)$-approximate algorithm.
In the following, we show that \cref{alg:degree=d} finds a desired solution in polynomial time.
By \cref{lem:degree=d:prob:A} and \cref{lem:degree=d:prob:B}, we know that the probability that a bad event occurs is at most $\frac{1}{2ed}$, where $d$ is the maximum degree of the dependency graph.
By \cref{lem:degree=d:dependency-graph}, we know $d\geq 1$.
Hence, we have $\frac{1}{2ed}\leq \frac{1}{e(d+1)}$.
Thus, we have $\Pr[C\text{ occurs}]\leq\frac{1}{e(d+1)}$ for any bad event $C\in\cA\cup\cB$.
We define $p:=\frac{1}{e(d+1)}$.
Therefore, we have that $e(d+1)\cdot p= e(d+1)\cdot \frac{1}{e(d+1)}=1\leq 1$.
Thus, the probability of defined bad events satisfies the condition of \cref{lem:sLLL}.
Since \cref{lem:gLLL} implies \cref{lem:sLLL} by setting $x_C=\frac{1}{d+1}$ for each bad event $C\in\cA\cup\cB$, then \cref{alg:degree=d} finds a solution that makes all bad events not occur in expected time at most $\sum_{C\in \cA\cup\cB}\frac{x_C}{1-x_C}=(1+\frac{1}{d})\cdot\abs{\cA\cup\cB}$ by \cref{lem:gLLL}.
Recall that $\abs{L}=n$ and $\abs{R}=m$.
By \cref{def:badevents}, we have $\abs{\cA}\leq n$ and $\abs{\cB}\leq \ceil{\frac{m}{\Delta}}\leq m$.
By \cref{def:dependency-graph}, we have $1\leq d \leq \Delta^2$.
Thus, $(1+\frac{1}{d})\cdot\abs{\cA\cup\cB}$ is $\poly(n,m)$.
Hence, the expected running time of \cref{alg:degree=d} is in $\poly(n)$, where $n$ stands for the number of vertices in the input bipartite graph.
\end{proof}

\section{Extensions}
\label{sec:weight}

In this section, we extend our results in \cref{sec:general} and \cref{sec:degree=d} to the weighted case, i.e., each set $j\in[m]$ in the input set system has a non-negative weight $w_j\geq 0$.
The goal is still to select $k$ sets from the collection of subsets.
Clearly, the following natural linear programming formulation \eqref{Feas-WLP} still has an integrality gap of $\Omega(\frac{\log n}{\log \log n})$ or $\Omega(\frac{\log\Delta}{\log \log\Delta})$, where $n$ is the number vertices in the input bipartite graph and $\Delta$ is the maximum degree.

\begin{align*}
     &&  & \tag{\text{Feas-WLP}} \label{Feas-WLP}\\
    &&\sum_{v\in N_G(u)}x_v\cdot w_v &\leq T, &\forall u\in L \\
    &&\sum_{v\in R}x_v &\geq k, & \\
    && 0\leq x_v &\leq 1, &\forall v\in R 
\end{align*}

We shall discuss the linear programming \eqref{Feas-WLP} a little bit more since all results in \cref{sec:weight:general} and \cref{sec:weight:degree=d} based on \eqref{Feas-WLP}.
We guess the value of the optimal solution whose range is from $0$ to $\max_{u\in L}\sum_{v\in N_{G}(u)}w_v$.
As a typical, we shall use the standard doubling technique to get a lower bound of the optimal solution.
We start with an upper bound of the optimal solution and consecutively decrease the upper bound until \eqref{Feas-WLP} does not admit a feasible solution.
See \cref{alg:doubling} for the formal description.

\begin{algorithm}[htb] 
\caption{Doubling Algorithm}
\label{alg:doubling}
\begin{algorithmic}[1]
\Require The linear programming formulation \eqref{Feas-WLP}.
\Ensure A lower bound $T^*$ of the optimal solution; A fractional solution $\bx^*$.
\State $\flag\leftarrow\true$; $T^*\leftarrow\max_{u\in L}\sum_{v\in N_{G}(u)}w_v$.
\While{$\flag=\true$}
\State $R'\leftarrow\set{v\in R\mid w_v\leq T^*}$.
\label{line:doubling:subgraph}
\State Set up \eqref{Feas-WLP}, and solve it with the guessed value $T^*$.
\If{\eqref{Feas-WLP} has a feasible solution}
\State Let $\bx^*$ be a feasible solution to \eqref{Feas-WLP}.
\State $T^*\leftarrow\frac{T^*}{2}$.
\Else
\State $\flag\leftarrow \false$.
\EndIf
\EndWhile
\State \Return $T^*\leftarrow 2T^*$ and $\bx^*$.
\end{algorithmic}
\end{algorithm} 

Let $T^*$ and $\bx^*=(x^*_u)_{u\in R}$ be the returned value of \cref{alg:doubling}.
It is not hard to see that \cref{alg:doubling} runs in polynomial time.
The returned $T^*$ is a good lower bound of the optimal solution; this can be captured by the following simple observation (\cref{obs:weight:T*}).

\begin{observation}
$1= T^*\leq 2\cdot\opt$\footnote{One can get a more accurate lower bound (e.g., $T^*\leq (1+\epsilon)\cdot\opt$) by the standard binary search technique; this shall lose more on the running time.}.   
\label{obs:weight:T*}
\end{observation}

\begin{proof}
Let $\lp^*$ be the minimum value of the guessed value $T$ such that \eqref{Feas-WLP} can have a feasible solution.
Clearly, $\lp^*$ is a lower bound of the optimal solution, i.e., $\lp^*\leq\opt$.
Note that \eqref{Feas-WLP} admits no feasible solution by using $\frac{T^*}{2}$, where $T^*$ is returned by \cref{alg:doubling}.
Thus, we have $\frac{T^*}{2}\leq \lp^* \leq \opt$.
Hence, $T^*\leq2\cdot\opt$.
After normalization, $T^*$ becomes $1$; see the following cutting and normalization.
\end{proof}

\paragraph{Cutting and Normalization.}
We first cut all large vertices, i.e., $\set{v\in R\mid w_v>T^*}$; this operation does not impact anything since $x^*_u=0$ for all $u\in \set{v\in R\mid w_v>T^*}$. 
We then normalize the weight of all vertices in $R$ based on the value of $T^*$, i.e., $w_v'\leftarrow\frac{w_v}{T^*}$ for all $v\in R$.
We remark that this produces an equivalent instance.
The normalization makes the value of $T^*$ be $1$, i.e., $\sum_{v\in N_{G}(u)}w_v'\cdot x^*_u\leq 1$ holds since (\rom{1}) $\sum_{v\in N_{G}(u)}w_u\cdot x^*_u\leq T^*$; (\rom{2}) $w_u'=\frac{w_u}{T^*}$ for all $u\in R$.
After normalization, the value of the optimal solution is also scaled by multiplying $\frac{1}{T^*}$.
In the remainder of this section, we use $\opt$ to denote the optimal solution to the normalized instance, and for each $u\in R$, $w_u$ is the normalized weight.
The cutting and normalization ensures the following two properties: (\rom{1}) $T^*=1$; (\rom{2}) $0\leq w_v\leq 1$ for all $v\in R$.
These two properties are crucial in the analysis of \cref{sec:weight:general} and \cref{sec:weight:degree=d}.

\subsection{General Graphs}
\label{sec:weight:general}

In this section, we show that a special case of the classical pipage rounding algorithm stated in \cref{alg:pipage} achieves $O(\frac{\log n}{\log \log n})$-approximation for \WKSS/ on general graphs (\cref{thm:weight:general}).

\begin{theorem}
Given any instance of \FKSS/ on general bipartite graphs, there is a randomized algorithm with running time $\poly(n)$ that returns a $O(\frac{\log n}{\log \log n})$-approximate solution with probability at least $1-\frac{1}{n}$, where $n$ is the number of vertices in the input bipartite graph.
\label{thm:weight:general}
\end{theorem}

Let $S$ be the set of vertices returned by \cref{alg:pipage}.
Recall that for each $v\in R$, $X_v\in\set{0,1}$ is a random variable that indicates whether $v$ is in $S$.
To prove \cref{thm:weight:general}, we need a new random variable $Y_v$.
For each vertex $u\in R$, we define a new random $Y_u\in\set{w_u,0}$ indicates the contribution of vertex $u$ in the solution $S$.
Note that $Y_v=w_v\cdot X_v$ for all $v\in R$.
The strong connection ensures that $Y_v$ also satisfies the marginal probability preservation and negative correlation properties; see \cref{obs:weight:general:Y_v}.

\begin{observation}
The random variables $\set{Y_u}_{u\in R}$ satisfy the following properties:
\begin{itemize}
    \item For each $u\in R$, $\Pr[Y_u=w_u]=x^*_u$;
    \item $\set{Y_u}_{u\in R}$ is negatively correlated.
\end{itemize}
\label{obs:weight:general:Y_v}   
\end{observation}

\begin{proof}
The first property holds since $\Pr[Y_u=w_u]=\Pr[X_u=1]=\Pr[u\in S]=x^*_u$ by \cref{lem:general:prob}.
The second property holds since for any $F\subseteq R$,
$$
\E\left[ \prod_{v\in F} Y_v \right] = \prod_{v\in F} w_v \cdot \E\left[ \prod_{v\in F} X_v \right] \leq \prod_{v\in F} w_v \prod_{v\in F}\E[X_v] = \prod_{v\in F}\E[Y_v].
$$
The inequality due to the fact that $\set{X_v}_{v\in R}$ is negatively correlated in \cref{lem:general:negative-correlation}.
\end{proof}

To prove \cref{thm:weight:general}, we need a weighted version of \cref{lem:general:ratio}; see \cref{lem:weight:general:ratio}.

\begin{lemma}
For each $u\in L$, $\Pr[\sum_{v\in N_G(u)\cap S}w_v\geq \frac{10\ln n}{\ln \ln n}\cdot T^*]\leq\frac{1}{n^2}$.
\label{lem:weight:general:ratio}
\end{lemma}

\begin{proof}
Note that $\Pr[Y_u=w_u]=x^*_u$ holds by \cref{obs:weight:general:Y_v}.
We define the random variable $Y:=\sum_{v\in N_{G}(u)}Y_v$.
We have the following upper bound of the expectation of the random variable $Y$: 
$$
\E[Y]=\sum_{v\in N_{G}(u)}w_v\cdot\Pr[Y_v=w_v]=\sum_{v\in N_{G}(u)}x^*_v\leq T^*=1.
$$
$T^*=1$ because we consider the normalized weight.
By \cref{obs:weight:general:Y_v}, $\set{Y_v}_{v\in R}$ is negatively correlated; thus, we use the Chernoff bound stated in \cref{def:chernoff}.
By choosing $(1+\delta)=\frac{10\ln n}{\ln \ln n}\geq 2$ for all $n> e$, we have:
\begin{align*}
&\Pr[Y\geq (1+\delta)\cdot T^*] \leq e^{-T^*(1+\delta)\ln(1+\delta)/4}  \tag{\cref{def:chernoff}} \\
\implies &\Pr[Y\geq \frac{10\ln n}{\ln \ln n}\cdot T^*] \leq e^{-T^*2\ln n} \tag{$\frac{10\ln n}{\ln\ln n}\ln\left( \frac{10 \ln n}{\ln \ln n} \right)\geq 8\ln n$} \\
\implies & \Pr[Y\geq \frac{10\ln n}{\ln \ln n}\cdot T^*] \leq e^{-2\ln n} = \frac{1}{n^2} \tag{$T^*= 1$}
\end{align*}
Thus, \cref{lem:weight:general:ratio} holds since $Y=\sum_{v\in N_{G}(u)\cap S}w_v$ for all $u\in L$.
\end{proof}

\begin{proof}[Proof of \cref{thm:weight:general}]
By \cref{lem:weight:general:ratio}, the returned solution $S$ have the following property:
$$
\Pr\left[\sum_{v\in N_G(u)\cap S}w_v\geq \frac{20\ln n}{\ln \ln n}\cdot \opt \right]\leq \Pr\left[\sum_{v\in N_G(u)\cap S}w_v\geq \frac{10\ln n}{\ln \ln n}\cdot T^*\right] \leq\frac{1}{n^2}, \text{ for all }u\in L.
$$
The first inequality is due to the fact that $T^*\leq 2\opt$ is stated in \cref{obs:weight:T*}.
Then, by union bound, we have:
$$
\Pr\left[\max_{u\in L}\sum_{v\in N_G(u)\cap S}w_v\geq \frac{20\ln n}{\ln \ln n}\cdot \opt \right] \leq \frac{1}{n}.
$$
Hence, \cref{alg:pipage} achieves $(\frac{20\ln n}{\ln \ln n})$-approximation with probability at least $1-\frac{1}{n}$.
\end{proof}

\subsection{Graphs with Bounded Degree}
\label{sec:weight:degree=d}

In this section, we show that \cref{alg:degree=d} is a $O(\log\Delta)$-approximate algorithm for \WKSS/ by slightly changing the definition of bad events (\cref{thm:weight:degree=d}).

\begin{theorem}
Given any instance of \WKSS/ on graphs with maximum degree $\Delta$, there is a randomized algorithm with running time $\poly(n)$ in the expectation that returns a $O(\log\Delta)$-approximate solution, where $n$ is the number of vertices in the input bipartite graph. 
\label{thm:weight:degree=d}
\end{theorem}

We start with the definition of bad events for \WKSS/ (\cref{def:badevents:weight}), where we only slightly change the definition of the performance bad events.

\begin{definition}[Bad Events for \WKSS/]
The bad event set $\cA\cup\cB$ consists of two types of bad events, where bad events $\cA$ and $\cB$ are called performance bad events (P-BE) and feasibility bad events (F-BE), respectively.
\begin{enumerate}[label=(S-SS*),leftmargin=*,align=left]
    \item[(P-BE)] For each vertex $u\in L'$, we have a bad event $A_u$ in $\cA$. Define $\cA_u\subseteq \cX$ as a subset of random variables, each of which corresponds to a vertex in $N_{G'}(u)$. We say $A_u$ occurs if:
    $$
    \sum_{i\in\cA_u} w_i\cdot X_i \geq 8 \cdot \ln(2e\Delta^2) \cdot (T^*+1).
    $$
    \item[(F-BE)] For each subgraph $D_j,j\in[\ell-1]$, we have a bad event in $\cB$. The last subgraph $D_{\ell}$ has a bad event if $\sum_{u\in D_{\ell}}(x^*_u+\frac{1}{\Delta})<1$; otherwise, $D_{\ell}$ does not have a bad event. For each bad event $B_j\in\cB$, define $\cB_j\subseteq \cX$ as a subset of random variables, each of which corresponds to a vertex in $D_j$. We say $B_j$ occurs if:
    $$
    \sum_{i\in\cB_j} X_i < \sum_{u\in D_j}x^*_u.
    $$
\end{enumerate}
\label{def:badevents:weight}
\end{definition}

It is not hard to see that \cref{lem:degree=d:dependency-graph} still holds since the dependency graph of bad events in \cref{def:badevents} is the same as \cref{def:badevents:weight}.
The rounding algorithm is also the same as \cref{alg:degree=d} by using the new definition of bad events above, i.e., replace line~\ref{line:degree=d:badevents} with \cref{def:badevents:weight}.
Similar to the analysis framework for the unweighted case, we start from the property of the vertex set $S_1$.

\begin{lemma}
$\max_{u\in L}\dis_u(S_1)\leq 4\ln(2e\Delta^2)(2\opt+1)$.
\label{lem:weight:degree=d:prop:S_1}
\end{lemma}

\begin{proof}
The first phase of \cref{alg:degree=d} is equivalent to a threshold rounding rounding by choosing $\theta:=\frac{1}{4\ln(2e\Delta^2)}-\frac{1}{\Delta}$.
Thus, for any $u\in L$, we have:
\begin{align*}
2\cdot \opt\stackrel{(\rom{1})}{\geq} T^* \geq \sum_{v\in N_G(u)}w_v\cdot x^*_v \stackrel{(\rom{2})}{\geq} \sum_{v\in S_1\cap N_G(u)} \theta\cdot w_v
&= \frac{\sum_{v\in N_G(u)\cap S_1}w_v}{4\ln(2e\Delta^2)}-\frac{\sum_{v\in N_G(u)\cap S_1}w_v}{\Delta} \\
&\stackrel{(\rom{3})}{=} \frac{\dis_u(S_1)}{4\ln(2e\Delta^2)}-\frac{\sum_{v\in N_G(u)\cap S_1}w_v}{\Delta} \\
&\stackrel{(\rom{4})}{\geq }\frac{\dis_u(S_1)}{4\ln(2e\Delta^2)}-\frac{\abs{N_G(u)\cap S_1}}{\Delta} \\
&\stackrel{(\rom{5})}{\geq }\frac{\dis_u(S_1)}{4\ln(2e\Delta^2)}-\frac{\Delta}{\Delta} \\
\end{align*}
The inequality (\rom{1}) is due to the fact that $T^*\leq 2\opt$ is stated in \cref{obs:weight:T*}.
The inequality (\rom{2}) is due to the fact that for all $v\in N_G(u)\cap S_1$, $x^*_v\geq \theta$.
The equality (\rom{3}) is due to the fact that $\dis_u(S_1)=\sum_{v\in N_G(u)\cap S_1}w_v$.
The inequality (\rom{4}) is due to the fact that $0\leq w_v\leq 1$ for all $v\in R$.
The inequality (\rom{5}) is due to the fact that $\abs{S_1\cap N_G(u)}\leq\abs{N_G(u)}\leq\Delta$.
\end{proof}

\cref{lem:weight:degree=d:prop:S_1} directly implies the ratio of \cref{alg:degree=d} in the first phase (\cref{obs:weight:degree=d:phase1}).

\begin{observation}
If \cref{alg:degree=d} does not enter the second phase, then the following two claims are true: (\rom{1}) $S=S_1$ is a feasible solution; (\rom{2}) $\alg\leq 4\cdot\ln(2e\Delta^2)\cdot (2\opt+1)$.  
\label{obs:weight:degree=d:phase1}
\end{observation}

Let $\fX$ be an assignment of random variables in $\cS$ that makes all bad events in $\cA\cup\cB$ not occur.
Let $S_2:=\set{v\in R'\mid X_v=1 \text{ in }\fX}$.
Note that we do not change the definition of the feasibility bad events.
Thus, \cref{lem:degree=d:phase2:demand} still holds.
By the definition of performance bad events, we have the following simple observation.

\begin{observation}
$\max_{u\in L'}\sum_{v\in N_{G'}(u)\cap S_2} w_v \leq 8\ln(2e\Delta^2)(2\opt+1)$.
\label{obs:weight:degree=d:prop:S_2}
\end{observation}
\begin{proof}
By \cref{def:badevents:weight}, if all bad events in $\cA$ do not occur, we have $\sum_{i\in\cA_u}w_i\cdot X_i<8\ln(2e\Delta^2)(T^*+1)$ for all $u\in L'$.
By \cref{obs:weight:T*}, we have $\sum_{v\in N_{G'}(u)\cap S_2} w_v \leq 8\ln(2e\Delta^2)(2\opt+1)$.
\end{proof}

By the same proof of \cref{lem:degree=d:prop:S_1+S_2}, we have the following observation.
Namely, the following two inequalities proves \cref{obs:weight:degree=d:prop:S_1+S_2}: (\rom{1}) $\max_{u\in L\setminus L'}\dis_u(S_1\cup S_2)\leq 4\ln(2e\Delta^2)(2\opt+1)$; (\rom{2}) $\max_{u\in L'}\dis_u(S_1\cup S_2)\leq 12\ln(2e\Delta^2)(2\opt+1)$.

\begin{observation}
$\max_{u\in L}\dis_u(S_1\cup S_2)\leq 12\ln(2e\Delta^2)(2\opt+1)$.
\label{obs:weight:degree=d:prop:S_1+S_2}
\end{observation}

Thus, we have the ratio of \cref{alg:degree=d} in the second (\cref{obs:weight:degree=d:phase2}), whose proof is the same as \cref{lem:degree=d:phase2} and \cref{lem:degree=d:phase3}.

\begin{observation}
If \cref{alg:degree=d} enters the second phase but does not enter the third phase, and $S_2$ is a solution that makes all bad events in $\cA\cup\cB$ do not occur, then the following two claims are true: (\rom{1}) $\abs{S_1\cup S_2}\geq k$; (\rom{2}) $\alg \leq 12\cdot\ln(2e\Delta^2)\cdot (2\opt+1)$.
\label{obs:weight:degree=d:phase2}    
\end{observation}

The proof of the ratio of \cref{alg:degree=d} in the third phases (\cref{obs:weight:degree=d:phase3}) is slightly different from \cref{lem:degree=d:phase3} since arbitrarily pick one vertex may not increase solution's value by $1$.

\begin{observation}
If \cref{alg:degree=d} enters the third phase and selects $v^*$ in the third phase.
Assume that $S_2$ is a solution that makes all bad events in $\cA\cup\cB$ do not occur, then the following two claims are true: (\rom{1}) $S_1\cup S_2 \cup\set{v^*}$ is a feasible solution; (\rom{2}) $\alg\leq 12\cdot\ln(2e\Delta^2)\cdot(3\opt+1)$.
\label{obs:weight:degree=d:phase3}
\end{observation}

\begin{proof}
The first claim of \cref{obs:weight:degree=d:phase3} is the same as \cref{lem:degree=d:phase3}.
By \cref{obs:weight:degree=d:prop:S_1+S_2}, we have:
\begin{align*}
\max_{u\in L}\dis_u(S_1\cup S_2 \cup \set{v^*})
&\leq \max_{u\in L}\dis_u(S_1\cup S_2) + \max_{u\in L}\dis_u(\set{v^*}) \\
&\leq 12\ln(2e\Delta^2)(2\opt+1) + w_{v^*} \tag{\cref{lem:degree=d:prop:S_1+S_2}} \\
&\leq 12\ln(2e\Delta^2)(2\opt+1) + T^* \tag{$w_{v^*}\leq T^*$} \\
&\leq 12\ln(2e\Delta^2)(2\opt+1) + 2\opt \tag{$T^*\leq 2\opt$} \\
&\leq 12\ln(2e\Delta^2)(3\opt+1) \tag{$2\leq 12\ln(2e\Delta^2)$}
\end{align*}

\end{proof}

Note that \cref{lem:degree=d:prob:B} still holds since we do not change the definition of the feasibility bad events.
In \cref{lem:weight:degree=2:prob:A}, we upper bound the probability that a performance bad event occurs.

\begin{lemma}
For all bad events $A_i\in\cA$, $\Pr[A_i\text{ occurs}]\leq\frac{1}{2ed}$, where $d$ is the maximum degree of the dependency graph $\dg(\cA\cup\cB)$.
\label{lem:weight:degree=2:prob:A}
\end{lemma}

\begin{proof}
Recall that for each vertex $v\in R'$, there is a random variable $X_v\in\set{0,1}$ that indicates whether the vertex $v$ is added to the solution $S_2$.  
For each vertex $v\in R'$, we define a new random variable $Y_v\in\set{w_v,0}$ that indicates the contribution of the vertex $v$ to the solution $S_2$.
Note that $\Pr[Y_v=w_v]=\Pr[X_v=1]=(x^*_v+\frac{1}{\Delta})\cdot 4\ln(2e\Delta^2)<1$.
We focus on an arbitrary bad event $A_i$, define $Y:=\sum_{v\in N_{G'}(i)}Y_v$.
We first upper bound the value of $\E[Y]$.
\begin{align*}
\E[Y]
&=4\cdot \ln(2e\Delta^2)\sum_{v\in N_{G'}(i)}w_v\cdot (x^*_v+\frac{1}{\Delta})
=4\cdot \ln(2e\Delta^2)\cdot\left(\frac{\sum_{v\in N_{G'}(i)}w_v}{\Delta}+\sum_{v\in N_{G'}(i)}w_v\cdot x^*_v\right)  \\
&\leq 4\cdot \ln(2e\Delta^2)\cdot\left(\frac{\abs{N_{G'}(i)}}{\Delta}+\sum_{v\in N_{G'}(i)}w_v\cdot x^*_v\right) \tag{$0\leq w_v\leq 1,\forall v\in R$} \\
&\leq 4\cdot \ln(2e\Delta^2)\cdot\left(1+\sum_{v\in N_{G'}(i)}w_v\cdot x^*_v\right) \tag{$\abs{N_{G'}(i)}\leq \abs{N_{G}(i)}\leq\Delta$} \\
&\leq 4\cdot \ln(2e\Delta^2)\cdot(1+T^*) \tag{$\sum_{v\in N_{G'}(i)}w_v x^*_v\leq \sum_{v\in N_{G}(i)}w_v x^*_v \leq T^*$} \\
\end{align*}
Now, we use the Chernoff bound above by setting $\delta=1$.
\begin{align*}
&\Pr[Y\geq (1+\delta)\cdot\hmu] \leq e^{-(1+\delta) \cdot \ln(1+\delta)\cdot \hmu /4} \tag{Chernoff Bound \cref{equ:chernoff:right}} \\
\implies & \Pr[Y\geq 2\cdot\hmu] \leq e^{-2\ln2\cdot\hmu/4} \tag{$\delta=1$} \\
\implies &\Pr[Y\geq 8\cdot\ln(2e\Delta^2)\cdot(1+T^*)] \leq e^{-\ln(2e\Delta^2)} = \left( \frac{1}{e} \right)^{\ln(2e\Delta^2)} \tag{$1+T^*=2>1$} \\
\implies &\Pr[Y\geq 8\cdot\ln(2e\Delta^2)\cdot(1+T^*)] \leq \frac{1}{2e\Delta^2} \leq \frac{1}{2ed} \tag{$\Delta^2\geq d$ in \cref{lem:degree=d:dependency-graph}}
\end{align*}
By \cref{def:badevents:weight}, we have:
$$
\Pr[A_i\text{ occurs}]=\Pr\left[\sum_{v\in\cA_i}w_v\cdot X_v\geq 8\ln(2e\Delta^2)(T^*+1)\right]=\Pr[Y\geq 8\ln(2e\Delta^2)(T^*+1)] \leq \frac{1}{2ed}.
$$
\end{proof}

\begin{proof}[Proof of \cref{thm:weight:degree=d}]
The running time \cref{alg:degree=d} is the same as the proof of \cref{thm:degree=d}.
By \cref{obs:weight:degree=d:phase1}, \cref{obs:weight:degree=d:phase2}, and \cref{obs:weight:degree=d:phase3}, we know that if \cref{alg:degree=d} returns a solution that makes all bad events defined in \cref{def:badevents:weight} not occur, then \cref{alg:degree=d} is a $12\ln(2e\Delta^2)(3\opt+1)$-approximate algorithm.    
\end{proof}

\section{Conclusion}

We study the fair k-set selection problem where we aim to select $k$ sets from a given set system such that the maximum occurrence times that an element appears in these $k$ selected sets are minimized.
Given a bipartite graph $G:=(L\cup R, E)$, our problem is equivalent to selecting $k$ vertices from $R$ such that the maximum number of selected neighbors of vertices in $L$ is minimized.
We demonstrate that the problem is $\NP$-hard when the maximum degree $\Delta=3$, and then we give a simple optimal algorithm for the $\Delta=2$ case.
We give two LP-rounding algorithms that are logarithmic approximate for general bipartite graphs and bipartite graphs with a maximum degree $\Delta$.

This work points out many interesting directions.
Firstly, it would be interesting to see whether the approximation can be further improved.
This may need a more involved LP formulation.
In particular, it would be interesting to see whether the problem admits a logarithmic lower bound.
Secondly, considering a more general case is interesting where we additionally require that the selected vertices form a base of some matroid.
The pipage rounding still works for this case, but it is not clear whether this general problem admits a $O(\log \Delta)$-approximate algorithm.
Lastly, it is also interesting to investigate other fairness criteria, such as the Nash social welfare or a $\ell_q$-norm objective.

\newpage
\clearpage
\bibliographystyle{plain}
\bibliography{references}

\newpage
\appendix
\section{Faster Algorithm for Unweighted Case and $\Delta=2$}
\label{app:sec:delta=2}

This section give a linear time algorithm for \FKSS/ instance on graphs with a maximum degree $2$ and all vertices in $R$ has a equal weight.
Observe that any \FKSS/ instance on graphs with a maximum degree $2$ has the optimal solution of either $1$ or $2$.
Thus, for this specific case, we only need an algorithm that is able to distinguish these two types of instances.
To this end, we shall solve the following problem: 

\begin{definition}[Maximum Vertex Set Problem (MVSP)]
Given a bipartite graph $G:=(L\cup R,E)$ with $\Delta \leq 2$, computing a subset $S\subseteq R$ with the maximum size such that the maximum disagreement over all vertices in $L$ is at most $1$.  
\label{def:degree=2:decision}
\end{definition}

Assume that we have such an algorithm, denoted by $\alg$ (\cref{alg:degree=2:decision}).
Let $S$ be the set of vertices returned by $\alg$, i.e., $S$ is the maximum size of vertices in $R$ so that the disagreement of all vertices in $L$ is smaller than $1$.
Then, if $\abs{S} \geq k$, we just arbitrarily pick $k$ vertices from $S$; so we find a solution with the maximum disagreement $1$.
If $\abs{S} < k$, then we know that the maximum disagreement of any solution is at least $2$; so we just arbitrarily pick $k$ vertices from $R$.
See \cref{alg:degree=2} for a formal definition.
Thus, if such an algorithm $\alg$ exists, \cref{thm:degree=2} holds by running \cref{alg:degree=2}.

\begin{theorem}
The problem of \FKSS/ is in $\PP$ on the bipartite graph $G:=(L\cup R, E)$ such that $\abs{N_{G}(u)}\leq 2$ for all $u\in L\cup R$.  
\label{thm:degree=2}
\end{theorem}

\begin{algorithm}[htb] 
\caption{Algorithm for \FKSS/ on graphs with $\Delta \leq 2$.}
\label{alg:degree=2}
\begin{algorithmic}[1]
\Require A bipartite graph $G=(L\cup R,E)$ with $\Delta \leq 2$; The demand $k\in\N_{\geq 1}$.
\Ensure A feasible solution $S\subseteq R$ with $\abs{S}=k$.
\State $S' \leftarrow \texttt{MaxVertexSet(G)}$. 
\Comment{call \cref{alg:degree=2:decision}.}
\IIf{$\abs{S'}\geq k$} $S\leftarrow$ any $k$ vertices from $S'$. \EndIIf
\IIf{$\abs{S'}< k$} $S\leftarrow$ any $k$ vertices from $R$. \EndIIf
\State \Return $S$.
\end{algorithmic}
\end{algorithm} 

Hence, in the following, we shall focus on the maximization problem stated in \cref{def:degree=2:decision}.
Observe that MVSP is actually a special case of the maximum weighted independent set on bipartite graphs with $\Delta \leq 2$, i.e., by assigning a big weight to each vertex in $L$ and a small weight to each vertex in $R$.
In this way, any optimal solution to the maximum independent set problem shall first include all vertices in $L$ and then try to include more vertices from $R$.
Hence, such an algorithm computes a subset of vertices in $R$ with the maximum size such that the disagreement of all vertices in $L$ is at most $1$.
It is well-known that the maximum weighted independent set on bipartite graphs can be solved in polynomial time by computing a minimum weighted vertex cover.
One can also use the algorithm for the maximum weighted independent set on graphs with bounded treewidth~\cite[Section 7.3.1]{DBLP:books/sp/CyganFKLMPPS15} since the graph with the maximum degree $2$ has treewidth $2$.
Computing a minimum weighted vertex cover on bipartite graphs needs the maximum flow techniques, and the latter algorithm on bounded treewidth is a dynamic programming algorithm.
Both these two techniques require a high running time.
However, for the problem in \cref{def:degree=2:decision}, a more efficient and simpler algorithm exists; see \cref{lem:degree=2:decision}.

\begin{algorithm}[htb] 
\caption{\texttt{MaxVertexSet($\cdot$)}}
\label{alg:degree=2:decision}
\begin{algorithmic}[1]
\Require A bipartite graph $G=(L\cup R,E)$ with $\Delta \leq 2$.
\Ensure A vertex set $S\subseteq R$.
\For{each connected component $C_i$ of $G$}
\If{$C_i$ is a path}
\State Pick the first vertex $v\in R$ and assign the index $\texttt{1}$ to it.
\State Assign $\set{2,3,\ldots,\ell}$ to the remaining vertices in $R$ one by one along the path.
\IIf{$\ell$ is an odd number} $S_i\leftarrow\set{1,3,\ldots,\ell}$ 
\EndIIf
\Comment{Path-2 in \cref{fig:degree=2}}
\IIf{$\ell$ is an even number} $S_i\leftarrow\set{1,3,\ldots,\ell-1}$
\EndIIf
\Comment{Path-1 in \cref{fig:degree=2}}
\EndIf
\If{$C$ is a cycle}
\State Pick an arbitrary vertex $v\in R$ and assign the index $1$ to it.
\State Assign $\set{2,3,\ldots,\ell}$ to the remaining vertices in $R$ one by one along the cycle.
\IIf{$\ell$ is an odd number} $S_i\leftarrow\set{1,3,\ldots,\ell-1}$ 
\EndIIf
\Comment{Cycle-1 in \cref{fig:degree=2}}
\IIf{$\ell$ is an even number} $S_i\leftarrow\set{1,3,\ldots,\ell-2}$ 
\EndIIf
\Comment{Cycle-2 in \cref{fig:degree=2}}
\EndIf
\EndFor
\State \Return $S\leftarrow \bigcup_{i}S_i$.
\end{algorithmic}
\end{algorithm} 

\begin{lemma}
\cref{alg:degree=2:decision} returns an optimal solution to the Maximum Vertex Set Problem stated in \cref{def:degree=2:decision} in $O(n)$ times, where $n$ is the number of vertices in the input bipartite graph.
\label{lem:degree=2:decision}
\end{lemma}

\begin{proof}
When the maximum degree $\Delta=2$, the bipartite graph consists of a batch of connected components, each of which is either a path or a cycle.
We focus on any connected component and assume that there are $\ell$ vertices from $R$ included in the connected component.
Since the graph is a bipartite graph, a vertex from $L$ must be adjacent to a vertex from $R$.
Based on the following simple observations, \cref{alg:degree=2} is optimal.
In the case where the connected component is a cycle and $\ell$ is an odd number, the optimal solution is at most $\floor{\frac{\ell}{2}}=\frac{\ell-1}{2}$; see Cycle-1 in \cref{fig:degree=2}.
In the case where the connected component is a cycle and $\ell$ is an even number, the optimal solution is at most $\frac{\ell}{2}$; see Cycle-2 in \cref{fig:degree=2}.
In the case where the connected component is a path and $\ell$ is an odd number, the optimal solution is at most $\ceil{\frac{\ell}{2}}=\frac{\ell+1}{2}$; see Path-2 in \cref{fig:degree=2}.
In the case where the connected component is a path and $\ell$ is an even number, the optimal solution is at most $\frac{\ell}{2}$; see Path-1 in \cref{fig:degree=2}.
\cref{alg:degree=2} is optimal since it finds a solution that matches the upper bounds of the optimal solution above.

\begin{figure}[htb]
    \centering
    \includegraphics[width=14cm]{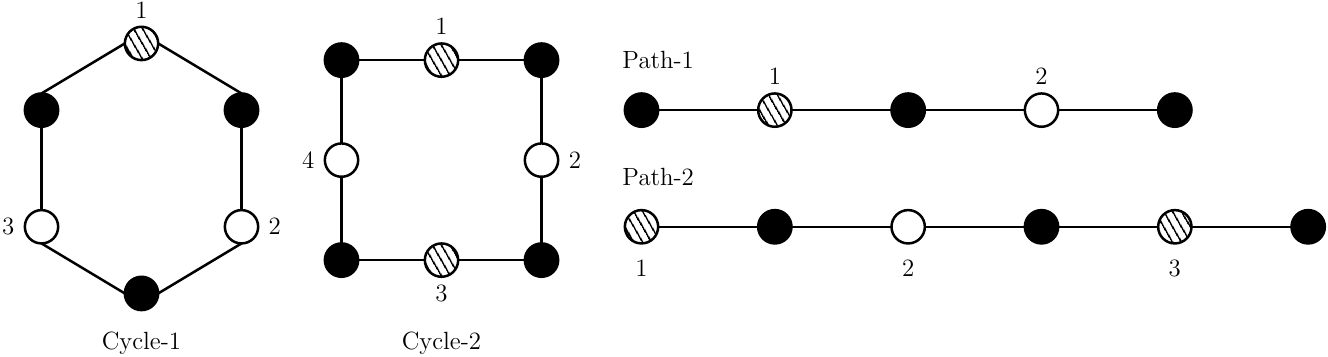}
    \caption{Illustration for \cref{alg:degree=2} and the proof of \cref{lem:degree=2:decision}. The black vertices are vertices in $L$, and all other vertices are in $R$. \cref{alg:degree=2} first assigns an index to each vertex from $R$ according to some fix direction. Depending on the parity of $\abs{R}$ and whether the connected component is a cycle, \cref{alg:degree=2} shall pick the different number of vertices from $R$ to satisfy the disagreement requirement. The shadow vertices are the vertices picked by \cref{alg:degree=2}.}
    \label{fig:degree=2}
\end{figure}

\end{proof}

\section{Proof of \cref{lem:general:pipage:pros}~\cite{DBLP:conf/focs/ChekuriVZ10}}
\label{app:lem:pipage}

In this section, we restate the proof of \cref{lem:general:pipage:pros} for completeness.
The initial proof of \cref{lem:general:pipage:pros} is given in~\cite{DBLP:conf/focs/ChekuriVZ10}.
The first part of \cref{lem:general:pipage:pros} is proven in \cref{lem:general:prob} and the second part is shown in \cref{lem:general:negative-correlation}.

We start with an important property of \cref{alg:pipage} (\cref{lem:general:pipage:expectation}), which is crucial to both the marginal probability preservation (\cref{lem:general:prob}) and the negative correlation property (\cref{lem:general:negative-correlation}).
Assume that \cref{alg:pipage} returns an integral solution after $t$ iterations.
Let $\xiu$ be a random variable that stands for the value of $x^*_u$ after the $i$-th iteration of \cref{alg:pipage}; so, $X_u^{(0)}=x^*_u$ with probability $1$ and $\xtu\in\set{0,1}$ is an integral random variable.
The correctness of \cref{lem:general:pipage:expectation} mainly depends on the right choice of the probability $p$ (line~\ref{line:pipage:prob} of \cref{alg:pipage}).

\begin{lemma}
For any iteration $i\in\set{0,1,\ldots,t-1}$ and any $u\in R$, we have $\E[\xiiu \mid \xiu]=\xiu$. 
\label{lem:general:pipage:expectation}
\end{lemma}

\begin{proof}
Consider any $i\in\set{0,1,\ldots,t-1}$ and $u\in R$, we distinguish two cases: Case (\rom{1}), $x^*_u$ is changed during the $i$-th iteration; Case (\rom{2}), $x^*_u$ is not changed.
In the second case, $\xiiu=\xiu$ with probability $1$; so, $\E[\xiiu \mid \xiu]=\xiu \cdot 1=\xiu$.
In the first case, there are two equivalent cases.
Either $\xiiu=\xiu-\delta_1$ with probability $p$ and $\xiiu=\xiu+\delta_2$ with probability $1-p$, or $\xiiu=\xiu+\delta_1$ with probability $p$ and $\xiiu=\xiu-\delta_2$ with probability $1-p$.
Thus, we have:
\begin{align*}
\E[\xiiu \mid \xiu] 
&= p\cdot (\xiu-\delta_1) + (1-p)\cdot (\xiu+\delta_2) \\
&= \frac{\delta_2}{\delta_1+\delta_2}\cdot (\xiu-\delta_1)+\frac{\delta_1}{\delta_1+\delta_2}\cdot (\xiu+\delta_2) \\
&=\frac{(\delta_1+\delta_2)\cdot \xiu}{\delta_1+\delta_2} \\
&=\xiu.
\end{align*}
For the case where $\xiiu=\xiu+\delta_1$ with probability $p$ and $\xiiu=\xiu-\delta_2$ with probability $1-p$, the calculation is the same as the above.
Thus, \cref{lem:general:pipage:expectation} holds.
\end{proof}

Recall that $\bx=(x_v^*)_{v\in R}$ is the optimal fraction solution.
Let $S\subseteq R$ be the set of vertices in $R$ that is returned by \cref{alg:pipage}.
For each vertex $v\in R$, let $X_v\in\set{0,1}$ be a random variable that indicates whether $v$ is in $S$, i.e., $X_v=1$ if $v\in S$; otherwise, $X_v=0$.
In the following, we shall frequently use the {\em law of total expectation}: for two random variables $X$ and $Y$, $\E[\E[X\mid Y]]=\E[X]$ assuming that all above expectations exist.

\begin{lemma}
For any vertex $u\in R$, we have $\Pr[u\in S] = x_u^*$.    
\label{lem:general:prob}
\end{lemma}
\begin{proof}
Consider any $u\in R$ and note that $\Pr[u\in S]=\Pr[\xtu=1]=\E[\xtu]$ since $\xtu\in\set{0,1}$ is a binary random variable.
Consider an arbitrary iteration $i\in\set{0,1,\ldots,t-1}$ and by \cref{lem:general:pipage:expectation}, we have $\E[\xiiu\mid\xiu]=\xiu$.
Taking expectations from both sides, we have $\E[\E[\xiiu\mid\xiu]]=\E[\xiu]$.
Applying the law of total expectation, we have $\E[\xiiu]=\E[\xiu]$.
Repeating above operations from $i=0$ to $i=t-1$, we have a chain of equations: $\E[\xtu]=\E[X_{u}^{(t-1)}]=\cdots=\E[X_{u}^{(0)}]$.
Note that $\E[X_{u}^{(0)}]=x^*_u$ since $X_{u}^{(0)}=x^*_u$ with probability $1$.
Thus, $\Pr[u\in S]=\E[\xtu]=x^*_u$ for any $u\in R$.
\end{proof}

\begin{lemma}
The collection $\set{X_1,\ldots,X_m}$ of random variables is negatively correlated.
\label{lem:general:negative-correlation}
\end{lemma}

\begin{proof}

Note that $X_u\in\set{0,1}$ is a binary random variable for all $u\in R$, thus $\E[X_u]=\Pr[X_u=1]=x^*_u$ for all $u\in R$ by \cref{lem:general:prob}.
Therefore, consider any vertex set $F\subseteq R$, we only need to show $\E[\prod_{u\in F}X_u]\leq \prod_{u\in F}x^*_u$.
To this end, we show that for any $i\in\set{0,1,\ldots,t-1}$, the following inequality holds:
\begin{equation}
\E\left[\prod_{u\in F}\xiiu\right]\leq\E\left[\prod_{u\in F}\xiu\right].    
\label{equ:negative-correlation:key}
\end{equation}
Applying above inequality from $i=0$ to $i=t-1$, we obtain a chain of inequalities:
$$
\E\left[\prod_{u\in F}X_u\right]=\E\left[\prod_{u\in F}X_{u}^{(t)}\right] \leq  \E\left[\prod_{u\in F}X_{u}^{(t-1)}\right] \leq \cdots \leq \E\left[\prod_{u\in F}X_{u}^{(0)}\right] = \prod_{u\in F}x^*_u.
$$
The first equation is due to the fact that $\xtu$ and $X_u$ are the same random variables.
The last equation is due to the fact that $X_{u}^{(0)}=x^*_u$ with probability $1$ for all $u\in R$.

In the following, we focus on proving \cref{equ:negative-correlation:key}.
To this end, we shall show the following conditional expectations hold:
\begin{equation}
\E\left[ \left.  \prod_{u\in F}\xiiu \right| \set{\xiv}_{v\in F} \right] \leq \prod_{u\in F}\xiu.
\label{equ:negative-correlation:key1}
\end{equation}
Taking expectations from both sides of \cref{equ:negative-correlation:key1} and applying the law of total expectation proves \cref{equ:negative-correlation:key}. 
Let $p,q\in R$ be the vertices such that the corresponding random variables are modified by \cref{alg:pipage} during the $(i+1)$-th iteration.
There are three cases: case (\rom{1}), $p,q\notin F$; case (\rom{2}), $p\in F,q\notin F$ or $p\notin F,q\in F$; case (\rom{3}), $p,q\in F$.
In the first case, $\xiiu=\xiu$ with probability $1$ for all $u\in F$.
Thus, conditioning on $\set{\xiv}_{v\in F}$, $\prod_{u\in F}\xiiu = \prod_{u\in F}\xiu$ with probability $1$ which implies \cref{equ:negative-correlation:key1}.
In the second case, assuming that $p\in F$ and $q\notin F$ and the other case is the same.
Then, we have $\xiiu=\xiu$ with probability $1$ for all $u\in F\setminus\set{p}$.
For the vertex $p$, we have $\E[\xiip\mid\xip]=\xip$ by \cref{lem:general:pipage:expectation}.
Thus, we have:
$$
\E\left[ \left.  \prod_{u\in F}\xiiu \right| \set{\xiv}_{v\in F} \right] = \prod_{u\in F\setminus\set{p}}\xiu \cdot \E\left[ \xiip \mid \xip \right]
= \prod_{u\in F\setminus\set{p}}\xiu \cdot \xip
= \prod_{u\in F}\xiu.
$$
For the last case, we only need to show the following inequality:
\begin{equation}
\E\left[ \xiip \cdot \xiiq \mid \xip,\xiq \right] \leq \xip \cdot \xiq.   
\label{equ:negative-correlation:key2}
\end{equation}
Based on \cref{equ:negative-correlation:key2}, we have:
\begin{align*}
\E\left[ \left.  \prod_{u\in F}\xiiu \right| \set{\xiv}_{v\in F} \right] 
&= \prod_{u\in F\setminus\set{p,q}}\xiu \cdot \E\left[ \xiip\cdot\xiiq \mid \xip,\xiq \right] \\
&\leq \prod_{u\in F\setminus\set{p,q}}\xiu \cdot \left( \xip\cdot \xiq \right) \\
&\leq \prod_{u\in F}\xiu.   
\end{align*}
The correctness of \cref{equ:negative-correlation:key2} mainly relies on another property of the pipage rounding:
$\xiip+\xiiq=\xip+\xiq$ holds with probability $1$.
Thus, we have 
\begin{align*}
&\E\left[(\xiip+\xiiq)^2\mid \xip,\xiq\right] = (\xip+\xiq)^2 \\
\implies & \E\left[ \left( \xiip \right)^2 + 2\cdot \left( \xiip \right) \cdot \left( \xiiq \right) + \left( \xiiq \right)^2 \mid \xip,\xiq \right] \\
&= \left( \xip \right)^2 + 2\cdot \left( \xip \right) \cdot \left( \xiq \right) + \left( \xiq \right)^2 \\
\implies &\E\left[ \left( \xiip \right)^2 \mid \xip \right] + \E\left[ 2\cdot \left( \xiip \right) \cdot \left( \xiiq \right) \mid \xip,\xiq \right] + \E\left[ \left( \xiiq \right)^2 \mid \xiq \right] \\
&= \left( \xip \right)^2 + 2\cdot \left( \xip \right) \cdot \left( \xiq \right) + \left( \xiq \right)^2
\end{align*}
By \cref{lem:general:pipage:expectation}, we have $\E[\xiiu\mid \xiu]=\xiu$ for all $u\in R$.
Thus, we have $\E[\xiiu-\xiiv\mid \xiu,\xiv]=\xiu-\xiv$ holds with probability $1$.
Hence, by using Jensen's inequality\footnote{If $X$ is a random variable and $\phi$ is a convex function, then $\phi(\E[X])\leq \E[\phi(X)]$.}, we have:
$$
\E\left[ (\xiip-\xiiq)^2 \mid \xip,\xiq \right] \geq \left(\E\left[ (\xiip-\xiiq) \mid \xip,\xiq \right] \right)^2=(\xip-\xiq)^2.
$$
Therefore, we have:
\begin{align*}
&\E\left[(\xiip-\xiiq)^2\mid \xip,\xiq\right] \geq (\xip-\xiq)^2 \\
\implies &\E\left[ \left( \xiip \right)^2 \mid \xip \right] - \E\left[ 2\cdot \left( \xiip \right) \cdot \left( \xiiq \right) \mid \xip,\xiq \right] + \E\left[ \left( \xiiq \right)^2 \mid \xiq \right] \\
&\geq \left( \xip \right)^2 - 2\cdot \left( \xip \right) \cdot \left( \xiq \right) + \left( \xiq \right)^2 \\
\implies &-\E\left[ \left( \xiip \right)^2 \mid \xip \right] + \E\left[ 2\cdot \left( \xiip \right) \cdot \left( \xiiq \right) \mid \xip,\xiq \right] - \E\left[ \left( \xiiq \right)^2 \mid \xiq \right] \\
&\leq -\left( \xip \right)^2 + 2\cdot \left( \xip \right) \cdot \left( \xiq \right) - \left( \xiq \right)^2
\end{align*}
Merge the above two long inequalities, we have
$$
4\cdot\E\left[ \left( \xiip \right) \cdot \left( \xiiq \right) \mid \xip,\xiq \right] \leq 4 \cdot \left( \xip \right) \cdot \left( \xiq \right).
$$
Hence, \cref{equ:negative-correlation:key2} holds.

\end{proof}

\end{document}